\documentclass[final,3p,times]{elsarticle}
\usepackage{amssymb}
\usepackage{amsmath}
\usepackage[version=4]{mhchem}
\usepackage{bm}
\usepackage{algorithm}
\usepackage{algpseudocode}
\usepackage{mathtools}
\usepackage{lineno,hyperref}
\usepackage{rotating}
\usepackage{subcaption}
\usepackage{makecell}%

\newcommand{\revone}[1]{\textcolor{black}{{#1}}}
\newcommand{\revtwo}[1]{\textcolor{black}{{#1}}}
\newcommand{\revonetwo}[1]{\textcolor{black}{{#1}}}
\newcommand{\revonerone}[1]{\textcolor{black}{{#1}}}
\newcommand{\revextrarone}[1]{\textcolor{black}{{#1}}}

\journal{TBD}

\biboptions{sort&compress}
\usepackage[dvipsnames]{xcolor}

\usepackage{lineno}

\usepackage{tikz}
\usepackage{etoolbox}
\newrobustcmd*{\myVtriangle}[2]{\tikz{\filldraw[draw=#1,fill=#2] (0cm,0.2cm) --
(0.2cm,0.2cm) -- (0.1cm,0cm) -- (0cm,0.2cm);}}
\newrobustcmd*{\mythickVtriangle}[2]{\tikz{\filldraw[line width=0.3mm,draw=#1,fill=#2] (0cm,0.2cm) --
(0.2cm,0.2cm) -- (0.1cm,0cm) -- (0cm,0.2cm);}}
\newrobustcmd*{\mythickErrorVtriangle}[2]{\tikz{\filldraw[line width=0.3mm,draw=#1,fill=#2] (-0.05cm,0.05cm) --
(0.05cm,0.05cm) -- (0cm,-0.05cm) -- (-0.05cm,0.05cm);  \draw[draw=#1] (0.0cm, -0.12cm) -- (0.0cm, 0.12cm) ; \draw[draw=#1] (-0.06cm, 0.12cm) -- (0.06cm, 0.12cm); \draw[draw=#1] (-0.06cm, -0.12cm) -- (0.06cm, -0.12cm)    }}
\newrobustcmd*{\mytriangle}[2]{\tikz{\filldraw[draw=#1,fill=#2] (0.0cm,0.0cm) --
(0.2cm,0cm) -- (0.1cm,0.2cm) -- (0cm,0cm);}}
\newrobustcmd*{\mysquare}[2]{\tikz{\draw[draw=#1,fill=#2] (0cm,0cm)
rectangle (0.2cm,0.2cm)}}
\newrobustcmd*{\mythicktriangle}[2]{\tikz{\filldraw[line width=0.3mm,draw=#1,fill=#2] (0.0cm,0cm) --
(0.2cm,0cm) -- (0.1cm,0.2cm) -- (0.0cm,0cm);}}
\newrobustcmd*{\mythicksquare}[2]{\tikz{\draw[line width=0.3mm,draw=#1,fill=#2] (0cm,0cm)
rectangle (0.2cm,0.2cm)}}
\newrobustcmd*{\mybarredtriangle}[2]{\tikz{\draw[draw=#1,fill=#2] (0,0) --
(0.2cm,0) -- (0.1cm,0.2cm) -- (0cm,0cm); \draw[draw=#1] (-0.1cm, 0.07cm) -- (0.3cm, 0.07cm)}}
\newrobustcmd*{\mythickbarredtriangle}[2]{\tikz{\draw[line width=0.3mm,draw=#1,fill=#2] (0,0) --
(0.2cm,0) -- (0.1cm,0.2cm) -- (0cm,0cm); \draw[draw=#1] (-0.1cm, 0.07cm) -- (0.3cm, 0.07cm)}}
\newrobustcmd*{\mybarredsquare}[2]{\tikz{\draw[draw=#1,fill=#2] (0,0)
rectangle (0.2cm,0.2cm); \draw[draw=#1] (-0.1cm, 0.1cm) -- (0.3cm, 0.1cm)}}
\newrobustcmd*{\mythickbarredsquare}[2]{\tikz{\draw[line width=0.3mm,draw=#1,fill=#2] (0,0)
rectangle (0.2cm,0.2cm); \draw[draw=#1] (-0.1cm, 0.1cm) -- (0.3cm, 0.1cm)}}
\newrobustcmd*{\mybarredcircle}[2]{\tikz{\draw[draw=#1,fill=#2] (0,0)
circle (0.1cm); \draw[draw=#1] (-0.2cm, 0.0cm) -- (0.2cm, 0.0cm)}}
\newrobustcmd*{\mythickbarredcircle}[2]{\tikz{\draw[line width=0.3mm,draw=#1,fill=#2] (0,0)
circle (0.1cm); \draw[draw=#1] (-0.2cm, 0.0cm) -- (0.2cm, 0.0cm)}}
\newrobustcmd*{\mythickErrorcircle}[2]{\tikz{\draw[line width=0.3mm,draw=#1,fill=#2] (0,0)
circle (0.06cm); \draw[draw=#1] (0.0cm, -0.12cm) -- (0.0cm, 0.12cm) ;   \draw[draw=#1] (-0.06cm, 0.12cm) -- (0.06cm, 0.12cm); \draw[draw=#1] (-0.06cm, -0.12cm) -- (0.06cm, -0.12cm)    }}
\newrobustcmd*{\mydashedline}[1]{\tikz{\draw[draw=#1] (-0.2cm, 0.2cm) -- (-0.1cm, 0.2cm); \draw[draw=#1] (-0.0cm, 0.2cm) -- (0.1cm, 0.2cm)}}
\newrobustcmd*{\mythickcross}[1]{\tikz{\draw[line width=0.3mm,draw=#1] (0,0) --
(0.2cm,0); \draw[line width=0.3mm,draw=#1] (0.1cm,-0.1cm) -- (0.1cm,0.1cm);}}
\newrobustcmd*{\mybarredcross}[1]{\tikz{\draw[line width=0.3mm,draw=#1] (0,0) --
(0.2cm,0); \draw[line width=0.3mm,draw=#1] (0.1cm,-0.1cm) -- (0.1cm,0.1cm); \draw[draw=#1] (-0.1cm,0) -- (0.3cm,0);}}
\newrobustcmd*{\myline}[1]{\tikz{\draw[draw=#1] (-0.15cm, 0.1cm) -- (0.15cm, 0.1cm);\draw[line width=0.3mm,draw=#1] (-0.0cm, 0.0cm);}}
\newrobustcmd*{\mythickline}[1]{\tikz{\draw[line width=0.3mm,draw=#1] (-0.15cm, 0.1cm) -- (0.15cm, 0.1cm);\draw[line width=0.3mm,draw=#1] (-0.0cm, 0.0cm);}}
\newrobustcmd*{\mythickdashedline}[1]{\tikz{\draw[line width=0.3mm,draw=#1] (-0.2, 0.1cm) -- (-0.1cm, 0.1cm); \draw[line width=0.3mm,draw=#1] (-0.0cm, 0.1cm) -- (0.1cm, 0.1cm); \draw[line width=0.3mm,draw=#1] (-0.0cm, 0.0cm);}}
\newrobustcmd*{\mythickdasheddottedline}[1]{\tikz{\draw[line width=0.3mm,draw=#1] (-0.22, 0.1cm) -- (-0.13cm, 0.1cm); \draw[line width=0.3mm,draw=#1] (-0.085cm, 0.1cm) -- (-0.055cm, 0.1cm); \draw[line width=0.3mm,draw=#1] (-0.01cm, 0.1cm) -- (0.08cm, 0.1cm); \draw[line width=0.3mm,draw=#1] (-0.0cm, 0.0cm);}}
\newrobustcmd*{\mycircle}[2]{\tikz{\draw[draw=#1,fill=#2] (0,0)
circle (0.1cm);}}
\newrobustcmd*{\mythickcircle}[2]{\tikz{\draw[line width=0.3mm,draw=#1,fill=#2] (0,0)
circle (0.1cm);}}
\newrobustcmd*{\mydot}[1]{\tikz{\draw[line width=0.3mm,draw=#1] (0,0)
circle (0.025cm);}}

\begin{document}

\begin{frontmatter}
\title{Bayesian calibration of bubble size dynamics applied to \ce{CO2} gas fermenters}

\author[NREL]{Malik Hassanaly}
\address[NREL]{Computational Science Center, National Renewable Energy Laboratory, Golden, CO 80401}
\author[NREL]{John M. Parra-Alvarez}
\author[NREL]{Mohammad J. Rahimi}
\author[NREL]{Federico Municchi}
\author[NREL]{Hariswaran Sitaraman}

\begin{abstract}
To accelerate the scale-up of gaseous \ce{CO2} fermentation reactors, computational models need to predict gas-to-liquid mass transfer which requires capturing the bubble size dynamics, i.e. bubble breakup and coalescence. However, the applicability of existing models beyond air-water mixtures remains to be established.
Here, an inverse modeling approach, accelerated with a neural network surrogate, calibrates the breakup and coalescence closure models, that are used in \revone{class methods for} population balance modeling (PBM). The calibration is performed based on experimental results obtained in a \ce{CO2}-air-water-coflowing bubble column reactor. Bayesian inference is used to account for noise in the experimental dataset and bias in the simulation results. To accurately capture gas holdup and interphase mass transfer, the results show that the breakage rate needs to be increased by one order of magnitude. The inferred model parameters are then used on a separate configuration and shown to also improve bubble size distribution predictions.
\end{abstract}

\end{frontmatter}






\section{Introduction}

\subsection{Motivation}
\label{sec:motivation}
In order to control global warming to levels lower than 2$^{\circ}$C, \ce{CO2} emissions need to be rapidly reduced (by 37\% in 2035 compared to 2019 emissions) \cite{lee2023ipcc}. To this end, demand reduction or decarbonization of carbon-heavy industrial sectors such as chemical manufacturing and aviation \cite{bergero2023pathways} is urgently needed. For specific industrial sectors such as aviation, there remain significant challenges to electrification \cite{schwab2021electrification}, and recent policies have instead incentivized the production of sustainable aviation fuels (SAF) \cite{shahriar2022current}. 

Given the widespread availability of gaseous \ce{CO2} through point sources and direct air capture, a promising pathway towards SAF production is to use \ce{CO2} as a carbon feedstock, instead of oil-based products, and several \ce{CO2}-based SAF production methods have been developed so far \cite{grim2023feasibility}. In particular, microbial action to produce valuable SAF intermediates from \ce{CO2}/syngas or \ce{CO2}/green-\ce{H2} mixtures is increasingly pursued \cite{tarraran2023high}, building upon successful microbial genetic engineering strategies developments \cite{straub2014selective}. Once the appropriate microbial strain is optimized in the laboratory, fuel production heavily relies on efficiently delivering and dissolving \ce{CO2} into the liquid medium in a gas fermentation reactor. The rate at which gaseous species transfer into the liquid phase, i.e. the interphase mass transfer rate, is a significant contributor towards the overall yield of the system.

The interphase mass transfer itself mainly depends on the Henry saturation concentration that varies directly with pressure \cite{pfeifer2021archaea}, the interphase slip velocity, and the size of the gas bubbles that determine an overall interfacial area \cite{yan2023mechanisms}. Multiphase computational fluid dynamics (CFD) simulations can facilitate the optimization and scale-up of such gas fermenters that utilize \ce{CO2}, but require careful representation of bubble sizes to be predictive and accurate.  Several methods have been proposed to account for variable bubble size. \revone{In particular, methods that use a population balance modeling (PBM) approach have been often used for multiphase flows \cite{mcgraw1997description, krepper2007inhomogeneous, lo1996application, marchisio2003quadrature}}. In the context of bubbly flows, PBM requires closure modeling to represent how bubbles \revone{coalesce and break up}. The focus of this work is to improve coalescence and breakup modeling with a specific focus on interphase mass transfer of \ce{CO2}, and thereby help accelerate and derisk scale-up of \ce{CO2} bioconversion.

\subsection{Bubble size dynamics model calibration}

Throughout the past decades, the choice of the coalescence and breakup models has been observed to significantly affect the numerical predictions for liquid-liquid dispersion \cite{gao2016simulation} and gas-liquid dispersion \cite{kalal2014modelling,wang2005population,li2024cfd}. The effect of the coalescence and breakup models was observed on bubble size distribution \cite{kalal2014modelling,wang2005population} and the Sauter mean bubble diameter \cite{kalal2014modelling} or droplet diameter \cite{gao2016simulation}. Those observations \revone{have led to the conclusion that models} need to be calibrated before being used \cite{kalal2014modelling,laakkonen2006validation,laakkonen2007modelling,singh2009population,li2024cfd}. \revtwo{Here, model calibration refers to optimizing empirical parameters to match experimental measurements available for a specific reactor. In particular, it differs from a validation procedure which does not involve parameter adjustment \cite{gel2023comparison}. In the present context, experimental measurements may be describing the bubble size distribution \cite{laakkonen2007modelling} or variables that depend on the bubble size \cite{deckwer1978comprehensive}.}

Several breakup and coalescence model calibration approaches have been undertaken in the past, mostly for liquid-liquid systems \cite{alopaeus2002simulation,ruiz2005determination,azizi2011turbulently,castellano2019using,sathyagal1995solution,mignard2006determination}, but were also extended to gas-liquid systems \cite{laakkonen2006validation,laakkonen2007modelling} which are of interest here. Those various efforts have focused on calibrating models in different regimes or against different datasets, and have found different sets of optimal model parameters which can vary by orders of magnitude \cite{azizi2011turbulently}. This observation highlights the importance of calibrating bubble size model dynamics against targeted experiments \cite{singh2009population,laakkonen2006validation,laakkonen2007modelling}, and also suggests that parameter calibration is a challenging task.

The first difficulty highlighted by \citet{maluta2021effect} during a deterministic calibration process is that the calibrated parameters \revtwo{can} compensate for numerical errors rather than address modeling deficiencies. The second difficulty identified is that there could exist multiple optimal parameter sets that can explain the same experimental observations \cite{maass2012determination}. Therefore, the solution of the calibration procedure should be probabilistic (rather than deterministic) to capture the ensemble of model parameters that explain experimental observations \cite{solsvik2014population}. The last difficulty relates to the appropriate choice of experimental datasets: depending on the experimental dataset used for calibration, different parameter values can be identified \cite{solsvik2014population}. As a side note, the issues in bubble size dynamics calibration tasks have led to the development of models that do not contain adjustable parameters \cite{becker2014development} or that \revone{aim to be generally} applicable \cite{rzehak2017unified,besagni2017computational}.  

To address the first difficulty, appropriate mesh convergence studies are needed \cite{maluta2021effect}. To address the second difficulty, a probabilistic calibration is needed. Currently, most breakup and coalescence model calibration tasks reported, provide deterministic optimized model parameters \cite{ruiz2005determination, alopaeus2002simulation, azizi2011turbulently, castellano2019using, laakkonen2006validation, singh2009population}. Only a handful of studies analyze the uncertainty in the calibrated model parameters, by using a local sensitivity analysis on the optimal parameter set \cite{sathyagal1995solution, mignard2006determination}. The advantage of such methods is that they are efficient in terms of the number of forward simulations - i.e. numerical simulations that predict experimental observations for a given parameter set -, but do not necessarily describe the entire distribution of parameter values. Bayesian inference methods have been devised to accurately characterize the parameters' distributions, but tend to require a large number of forward simulations~\cite{tierney1994markov,roberts1994simple,gelman1997weak}. Bayesian inference can be efficiently used when the experimental observations can be simulated with fast physics models \cite{braman2013bayesian,hassanaly2021surface,bell2019bayesian}, or \revtwo{fast surrogate models \cite{hassanaly2023pinn2,smith2021Atikokan,gel2023comparison}.}\revtwo{ As a side note, sensitivity analysis and calibration can be complementary when one wants to reduce the set of parameters to calibrate \cite{vaidheeswaran2021sensitivity,gao2021global}.} To address the third difficulty, combining multiple experimental datasets in the calibration procedure is necessary. When calibrating coalescence and breakup models, experimental datasets are usually combined by minimizing an error fit metric averaged across all the experimental data \cite{castellano2019using,solsvik2014population}. However, experimental data may be subject to different levels of noise, and experimental datasets should not be equally weighted in the parameter fitting procedure. Including experimental noise in the calibration procedure is however straightforward using a Bayesian calibration approach \cite{braman2013bayesian}. 

In the context of sustainable fuel development, there is evidence that interphase mass transfer can be strongly affected by the nature of the broth used for fermentation \cite{volger2023bubbles} and surfactants \cite{mcclure2015impact,painmanakul2005effect}. Those effects are only partially understood and not specifically modeled in general. Throughout this work, we approach this problem through the lens of model correction. Given the sensitivity of computational models to bubble dynamics model parameters, we propose to use bubble dynamics models as a knob that allows to improve the prediction of specific quantities of interest like gas holdup and interphase mass transfer.

The novel contributions of this work are
\begin{itemize}
    \item Coalescence and breakup models are calibrated against experimental data of a coflowing bubble column reactor using \ce{CO2} in the gas phase. Compared to previous work, the models are calibrated against gas holdup and interphase mass transfer, which are critical parameters for gas fermentation applications. 
    \item A Bayesian inference approach is used to calibrate efficiency factors for coalescence and breakup rates, thereby appropriately estimating confidence intervals on the efficiency factors. The Bayesian inference procedure is made tractable by constructing a data-based surrogate model.
    \item Four experimental \revtwo{gas-liquid} datasets are combined by taking into account \revtwo{the uncertainties due to missing physics in the numerical models}. This \revtwo{approach is used to determine the best bubble dynamics models} irrespective of the \revtwo{choice of the parameter value}.
    \item We show that the model parameters calibrated are also beneficial for predicting the bubble size distribution of a different reactor type \revone{(stirred-tank reactor)}.
\end{itemize}

In Sec.~\ref{sec:numerical} the numerical modeling approach is described. In Sec.~\ref{sec:deckwer}, the performances of the non-calibrated models are compared to the experimental dataset of interest \cite{deckwer1978comprehensive}.  The model calibration problem and the calibration results are shown in Sec.~\ref{sec:model_cal}. Conclusions are provided in Sec.~\ref{sec:conclusions}.

\section{Numerical method}
\label{sec:numerical}
\subsection{Multiphase model}

The numerical simulations are conducted with a multiphase solver implemented in OpenFOAM \cite{weller1998} that has been extensively used and validated in the past \cite{rahimi2018,lehnigk2022open,sitaraman2023reacting}. The multiphase model is implemented via the Euler-Euler method where gas and liquid are treated as continuous interpenetrating phases. Gas and liquid volume fractions are transported according to
\begin{equation}
    \label{eq:phaseCont}
    \frac{\partial (\alpha_j \rho_j)}{\partial t} + \nabla \cdot (\alpha_j \rho_j u_j) = \sum_{k\neq j} \dot{m}_{kj},
\end{equation}

where $j$ is the phase index (gas or liquid), $\alpha_j$ is the volume fraction of the phase $j$, $\rho_j$ is the density of phase $j$, $u_j$ is the velocity vector of phase $j$. The right-hand side of this equation includes the sum of interphase mass transfer terms where $\dot{m}_{kj}$ is the mass transfer rate from phase $k$ to phase $j$. To model the relative motion of each phase, a phase-momentum transport equation is solved, which takes the form

\begin{equation}
    \label{eq:phaseMom}
    \frac{\partial (\alpha_j \rho_j u_j)}{\partial t} + \nabla \cdot (\alpha_j \rho_j u_j u_j) = \nabla \cdot (\alpha_j \overline{\tau}) - \alpha_j \nabla p + \alpha_j \rho_j g +\sum_{k\neq j} D_{kj} + M^{m}_j + F_j, 
\end{equation}

In equation~\eqref{eq:phaseMom}, coupling between phase volume fraction and momentum needs to be enforced, including across phases, where drag forces depend on the relative velocities of each phase. Here, $\overline{\tau}$ is the stress tensor which includes Reynolds stress and viscous (molecular and turbulent) stress (including turbulent viscosity) for the phase $j$; $D_{kj}$ is the drag force exerted by phase $k$ on phase $j$ that depends on the drag coefficient associated with the dispersed phase and associated volume fraction, and $F_j$ contains interfacial forces acting on phase $j$ which includes \revone{lift}, wall-lubrication, virtual-mass and turbulent-dispersion forces, as described in our previous work \cite{rahimi2018}. In addition, momentum can be added to the system due to mass transfer; mass entering/leaving a specific phase. The term $M^{m}_j$ accounts for the added momentum:
\begin{equation}
    \label{eq:Mom_mass}
    M^{m}_j = \dot{m}_{j,in} u_k - \dot{m}_{j,out} u_j
\end{equation}
where $\dot{m}$ is the mass transfer rate from/to phase $j$, $u_j$ is the velocity of phase $j$ and $u_k$ is the velocity of phase $k$ \cite{lehnigk2022open}. 
A partial elimination method formulated by \citet{passalacqua2011implementation} is used here to ensure appropriate coupling between the phases. A shared pressure equation is solved once for all the phases to determine pressure while enforcing an aggregate divergence constraint on phase velocity. At the wall, a no-slip boundary condition was used for the liquid phase and a slip boundary condition was used for the gas phase. 

An energy equation is solved to account for differences in inlet phase temperatures ($287.15$K for the liquid phase and $273.15$K for the gas phase as mentioned in Sec.~\ref{sec:exp_dat}). A sensible energy transport equation for phase $j$ is written as:
\begin{equation}
\label{eq:energy_phase}
    \frac{\partial \rho_j \alpha_j E_j}{\partial t} + \nabla \cdot (\rho_j \alpha_j u_j E_j) = \nabla \cdot (\alpha_j \kappa_{j} \nabla T_j) + \dot{Q},
\end{equation}

where $E_j$ is the sensible energy enthalpy of the phase $j$, $T_j$ is the temperature of the phase $j$, $\kappa_{j}$ is the effective thermal diffusivity of the phase $j$ (including molecular and turbulent contributions) and $\dot{Q}$ is the heat exchange due to differences in temperature. Heat exchange through the interface is driven by temperature differences between phases. In particular the last term of equation~\eqref{eq:energy_phase} can be written as:
\begin{equation}
    \label{Q_temperature}
    \dot{Q} = h_{jl}(T_f - T_j)
\end{equation}
where $h_{jl}$ is the heat transfer coefficient of species $l$ in phase $j$, $T_j$ is the temperature of phase $j$ and $T_f$ is the temperature at the interface. The temperature at the interface is computed based on the assumption that the rate of heat transfer must be equal to the latent heat $\lambda_j$ at the interface between phases $j$ and $k$
\begin{equation}
    \label{temp_interface}
    h_{jl}(T_j - T_f) + h_{kl}(T_k - T_f) = \dot{m}_{j,in} \lambda_j.
\end{equation}
Although the energy equation is modeled to avoid unnecessary assumptions, the effect of heat exchange is likely negligible due to the narrow temperature difference between the liquid and the gas phase. Using the momentum and phase fraction, species mass fractions are transported as 

\begin{equation}
    \label{eq:massFrac}
    \frac{\partial (\rho_j \alpha_j Y_{jl})}{\partial t} + \nabla \cdot (\rho_j \alpha_j Y_{jl} u_j) = \nabla \cdot (\rho_j \alpha_j D_{jl} \nabla Y_{jl}) + S_{jl}
\end{equation}

where $Y_{jl}$ is the $l^{th}$ species mass fraction in phase $j$, $D_{jl}$ is the diffusivity of the $l^{th}$ species in phase $j$ and $S_{jl}$ are source terms due to interphase mass transfer. 

The aforementioned conservation equations are closed through models for interphase physics and turbulent stress. The interphase drag force is obtained by using the Grace model \cite{grace1976shapes} and transverse lift from the model by \citet{tomiyama2002transverse}. Wall lubrication forces are computed using the model by \citet{antal1991analysis} and turbulent dispersion uses the model of \citet{burns2004favre}. Interphase mass transfer of species is modeled by using the mass transfer coefficient obtained from Higbie correlation \cite{higbie1935rate}. The interphase mass transfer of a species also depends on the local saturation concentration obtained from Henry's constant and the local gas phase concentration of the species. Therefore, the local gas density, which is determined by the local static pressure, plays a significant role in interphase mass transfer. Turbulent stress is \revone{obtained from a mixture} $k-\varepsilon$ RANS closure model \cite{behzadi2004modelling} for both the liquid and the gas phases, \revone{wherein the bubble-induced turbulence model of Lahey~\cite{lahey2005simulation} is used.} The wall boundary conditions for $k$ and $\varepsilon$ use generalized wall functions \cite{popovac2007compound}. \revone{Since $\varepsilon$ is used for modeling bubble breakage induced by turbulence \cite{laakkonen2007modelling}, the modeling procedure of $\varepsilon$ plays an essential role in this work.}

\subsection{Bubble size models}

Compared to our previous works \cite{rahimi2018, sitaraman2023reacting}, the bubble size distribution (BSD) in this work is not assumed to be a Dirac $\delta$-distribution (constant bubble size), but is directly modeled using a population balance equation. In separate studies, several authors have noted the effect of BSD on global hydrodynamics of bubble column reactors \cite{buwa2002dynamics,diaz2008numerical,colella1999study,lucas2019influence} and on species mass transfer \cite{ramezani2012improved} which is the main focus here. 

The population balance model uses a method of classes implementation as described by \citet{lehnigk2022open} where the bubble size distribution is represented through a discretization of the number density function (NDF). This approach requires discretizing an internal coordinate (here the bubble size coordinate) and the spatial coordinates.
The fundamental governing equation of the NDF can be derived through a finite volume analysis \cite{lehnigk2022open} and leads to

\begin{equation}
    \label{eq:fi}
    \frac{\partial n_v}{\partial t} + \nabla \cdot (u n_v) = h_v, 
\end{equation}

where $u$ is the phase velocity, $t$ is time, $n_v$ is the number density of bubbles of size $v$ and the source term is $h_v =  B_{\rm b} - B_{\rm d} +  C_{\rm b} -  C_{\rm d} \revone{-} D + \dot{N_v} $, where  $B_{\rm b}$ (resp. $C_{\rm b}$) is the bubble birth contribution from bubble breakup (resp. coalescence), $B_{\rm d}$ (resp. $C_{\rm d}$) is the bubble death contribution from bubble breakup (resp. coalescence), $D$ is the drift term that is due to changes of bubble volume \revone{arising from pressure or mass transfer induced density changes}, and $\dot{N_v}$ is the bubble nucleation source term. The full expression of $h_v$ is available elsewhere (see Eq.2 in Ref.~\cite{lehnigk2022open}) and only specific modeling choices are described hereafter. The breakup and coalescence source terms depend on closure models for coalescence and breakup frequencies (also referred to as rates or kernels). For instance, the birth by breakup effect of bubbles of volume $v'$ to bubbles of volume $v$ ($v<v'$) is represented in $h_v$ as

\begin{equation}
   B_{\rm b} = \int_{v}^{\infty} B_{v'} \beta_{v,v'} n_{v'} dv',
\end{equation}

where $B_{v'}$ is the breakup frequency of bubbles of volume $v'$, and $\beta_{v,v'}$ is the daughter size distribution that describes how bubbles are redistributed within the internal coordinate (the binned bubble size). Similarly, the birth by coalescence effect of bubbles with volume $v'$ and $v-v'$ resulting in a bubble of volume $v$ is represented in $h_v$ as 

\begin{equation}
   C_{\rm b} = \frac{1}{2}\int_{0}^{v} n_v' n_{v-v'} C_{v',v-v'} dv',
\end{equation}

where $C_{v',v-v'}$ is the frequency of coalescence for bubbles of size $v'$ and $v-v'$.

A noteworthy feature of the implementation of \citet{lehnigk2022open} is that it includes redistribution schemes that conserve the 0$^{th}$ and 1$^{st}$ order moments of the NDF \cite{kumar1996solution, liao2018discrete}. The coalescence kernel chosen is that of \citet{lehr2002bubble} using a critical coalescence velocity of $0.08$m/s and a maximum packing density of $0.6$. In this model, the coalescence frequency $C_{v',v-v'}$ depends on turbulent fluctuations, and coalescence occurs if bubbles collide with a velocity greater than the critical velocity aforementioned. Two breakup models are investigated: a \textit{global breakup model} described in \citet{laakkonen2007modelling} and a \textit{binary breakup model} described in \citet{lehr2002bubble} which is common in bubble column reactor modeling \cite{kostoglou2005toward}. In the binary breakup approach, a bubble necessarily breaks up into two daughter bubbles from different size groups (or bins). In the global breakup approach, a bubble may break up into more than two bubbles \cite{laakkonen2007modelling}. The number of ``daughter bubbles" is controlled by the daughter size distribution.
The default implementation of these models \cite{lehnigk2022open,li2024cfd} in OpenFOAM v9 \footnote{https://github.com/OpenFOAM/OpenFOAM-9} is used unless stated otherwise.

\section{Validation of the base model}
\label{sec:deckwer}

In this section, the experimental case simulated is described. A base simulation is compared to the experimental results and the effect of numerical errors and modeling errors is discussed. 

\subsection{Experimental data}
\label{sec:exp_dat}

Two experimental cases are used for model comparison here and documented in \citet{deckwer1978comprehensive} as experiments 17 and 19. The system consists of a coflowing bubble column reactor schematically illustrated in Fig.~\ref{fig:CO2} (left). Both cases differ in terms of their superficial gas velocity and inlet gas composition. These cases were chosen given that they were also used in other numerical analyses such as by \citet{hissanaga2020mass} and by \citet{ngu2022spatio}. In Exp.~17 the inlet gas velocity is $3.42 \rm{cm/s}$ and the inlet gas \ce{CO2} mole fraction is $0.673$. In Exp.\ 19, the inlet gas velocity is $4.63\rm{cm/s}$ and the inlet gas \ce{CO2} mole fraction is $0.478$. The liquid phase is made of pure water at the inlet. Air is assumed to be pure \ce{N2} for simplification. The gas phase inlet temperature is set to $273.15$K consistently with \citet{deckwer1978comprehensive} and the liquid phase (tap water) inlet temperature is assumed to be $287.15$K.

\subsection{Numerical details}
\label{sec:numerics}

\begin{figure}[h]
    \centering
    \includegraphics[width=0.5\textwidth]{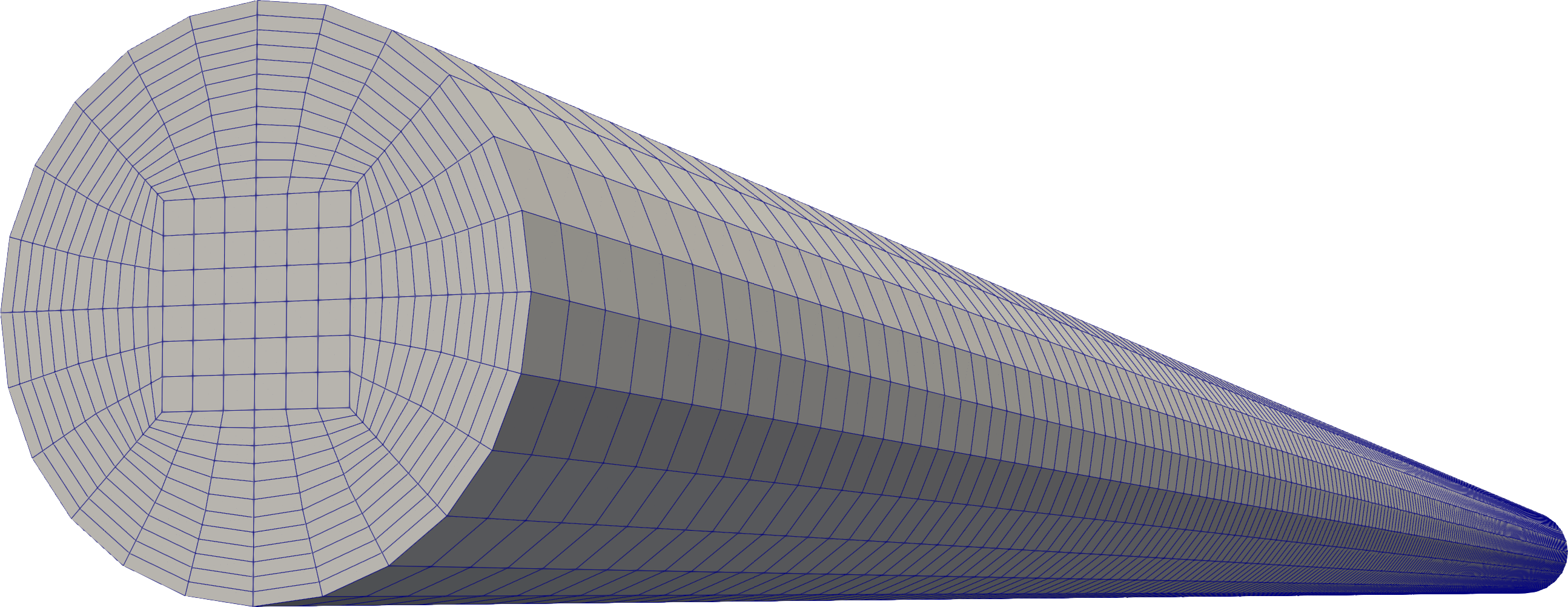}
    \caption{Illustration of the pillow case mesh in the radial and azimuthal directions.}
    \label{fig:pillow}
\end{figure}

A hexahedral mesh was used to simulate these experiments using 180 mesh points of size 2.4 cm each in the vertical direction. A ``pillow case" meshing (illustrated in Fig.~\ref{fig:pillow}) was used in the radial and azimuthal direction with 15 mesh points through the radial direction and 24 points in the azimuthal direction. In total, 64,800 cells were used. 

The internal coordinate (bubble diameter) was discretized with 10 classes (bins), uniformly distributed between $1.66~\rm{mm}$ and $4.36~\rm{mm}$. In separate numerical investigations \revone{\cite{ngu2022spatio,rzehak2016euler}}, the bubble size was assumed constant and equal \revone{or near} $2.86~\rm{mm}$ consistently with experimental data \cite{deckwer1978comprehensive}. \revtwo{In \ref{app:effectBSD}, a discussion about the effect of the bubble size range on the calibration results is provided.}  At the initial time, the NDF is initialized at 0 for all the classes except the one centered at $2.86~\rm{mm}$. The gas phase is initialized as pure \ce{N2} and the liquid phase is initialized as pure water. Hereafter, the global breakup model and binary breakup model as described in Sec.~\ref{sec:numerical} are successively used. Simulations of both Exp.~17 and Exp.~19 use the same numerical discretization. The liquid diffusivity of \ce{CO2} was set to $1.4663\times 10^{-9} $ m$^2$/s and the Henry's constant of \ce{CO2} was set to $0.0192$ mg/L/Pa which are the same values used by \citet{ngu2022spatio}. The simulations were run for $400~\rm{s}$ which was found to be sufficient to reach steady state in terms of gas holdup and species concentration profiles. Each simulation required about 10 hours of wall clock time on 36 cores of Dual Intel Xeon Gold Skylake 6154 3.0 GHz processors. The timestep is dynamically adjusted based on the maximal Courant number (set to 0.5). The minimum timestep is set to $0.1~\rm{ms}$ and is only used for the first timestep before ramping up to about $2~\rm{ms}$. Inlet boundary conditions for $k$ and $\varepsilon$ based on freestream turbulence were found to lead to steep wall-normal gradients due to the wall functions used, and could lead to unphysical oscillations in $\alpha_j$. The boundary conditions were adjusted to eliminate these oscillations without inducing appreciable differences in the quantities of interest (see \ref{app:bcke}).   

\begin{figure}[h!]
    \centering
    \includegraphics[width=0.55\textwidth]{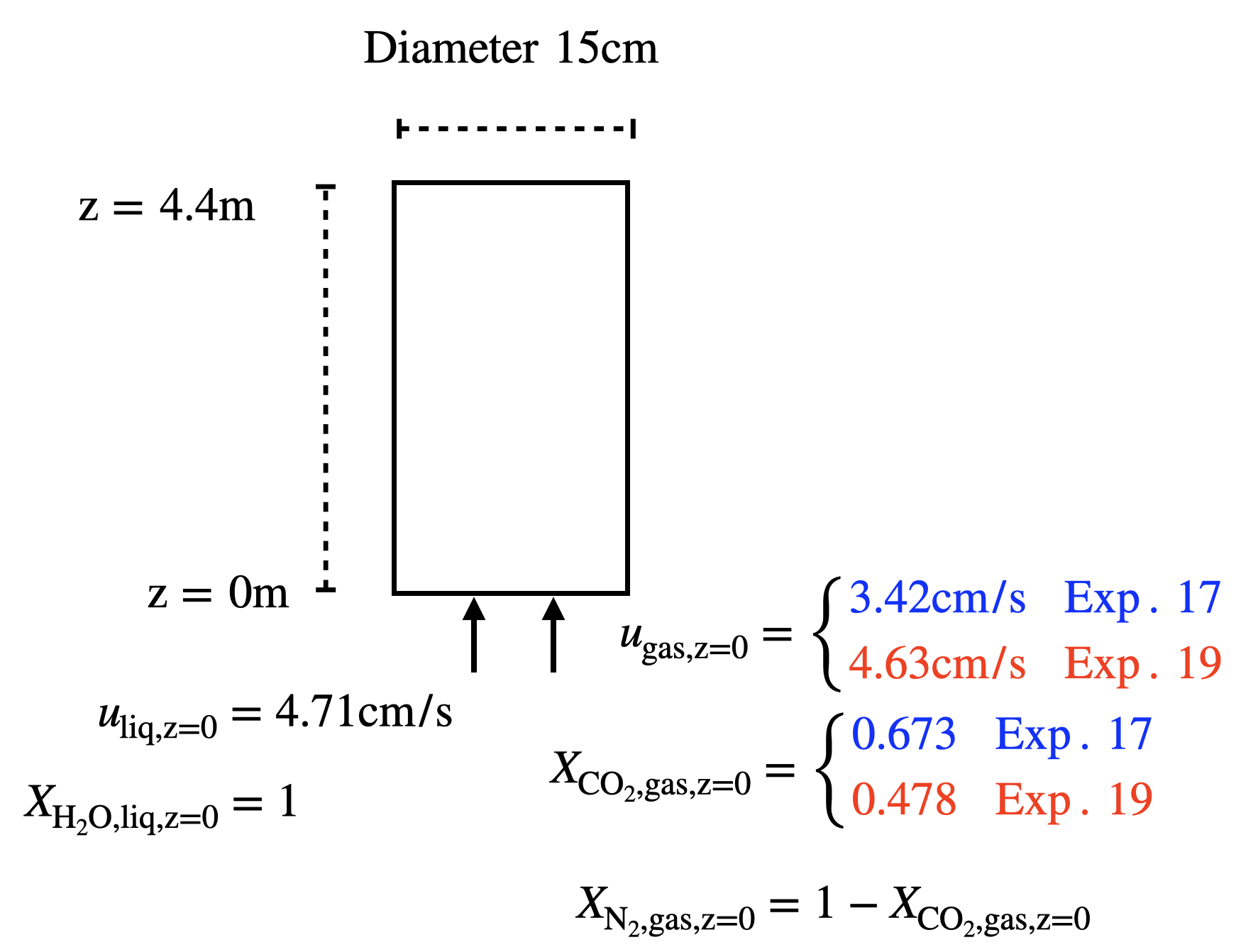}
    \includegraphics[width=0.22\textwidth]{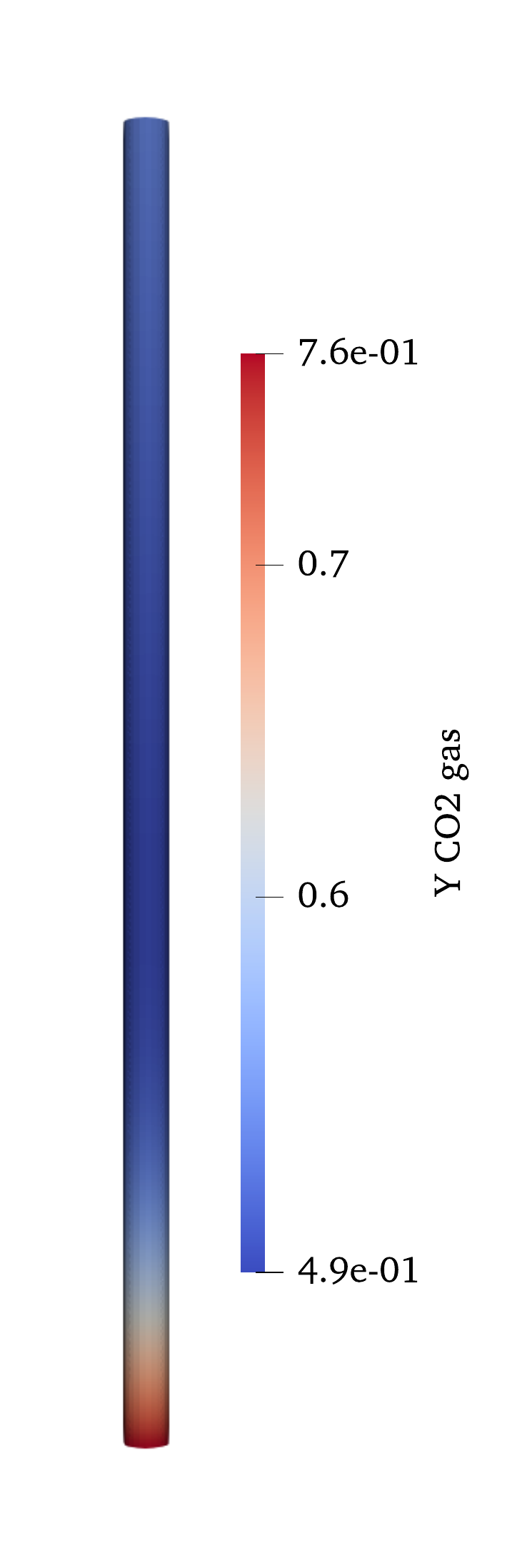}
    \includegraphics[width=0.22\textwidth]{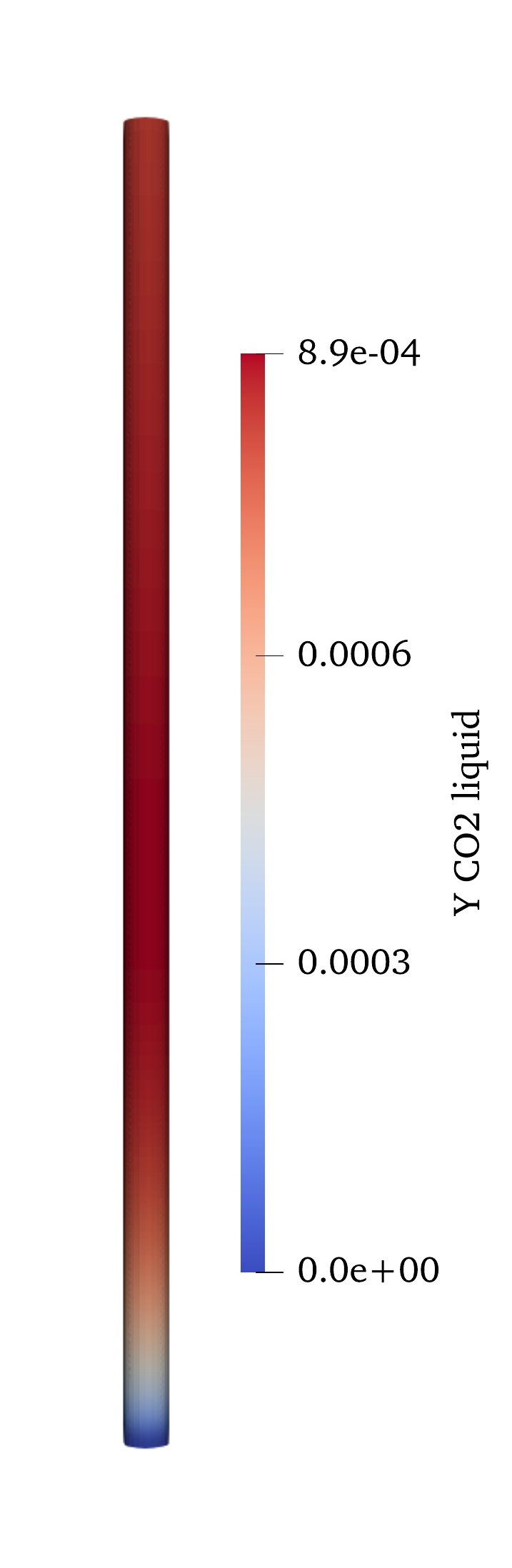}
    \caption{\revone{Left: schematic illustration of the case simulated with inlet conditions for Exp.~17 (blue) and Exp.~19 (red). Right: \revonerone{volumetric contours of \ce{CO2} mass fraction} at steady state for Exp.~17 of \citet{deckwer1978comprehensive} in the gas phase (left) and the liquid phase (right).}}
    \label{fig:CO2}
\end{figure}

Figure~\ref{fig:CO2} (right) shows the \revonerone{volumetric} \ce{CO2} mass fraction contour in the gas phase and the liquid phase for Exp.~17 which illustrates how the interphase mass transfer occurs in the coflowing bubble column. The gas phase is gradually depleted of \ce{CO2} with height (z in Fig.~\ref{fig:CO2}) while the liquid phase \ce{CO2} concentration increases. 

\subsection{Comparison to experiments}
\label{sec:comp_exp}

\begin{figure}[h]
    \centering
    \includegraphics[width=0.4\textwidth]{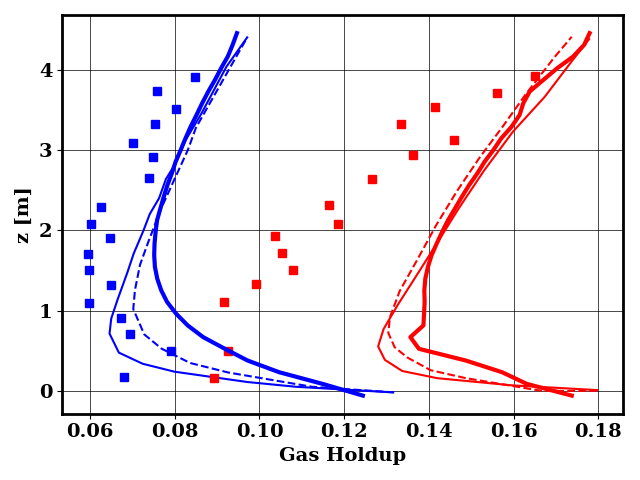}
    \includegraphics[width=0.4\textwidth]{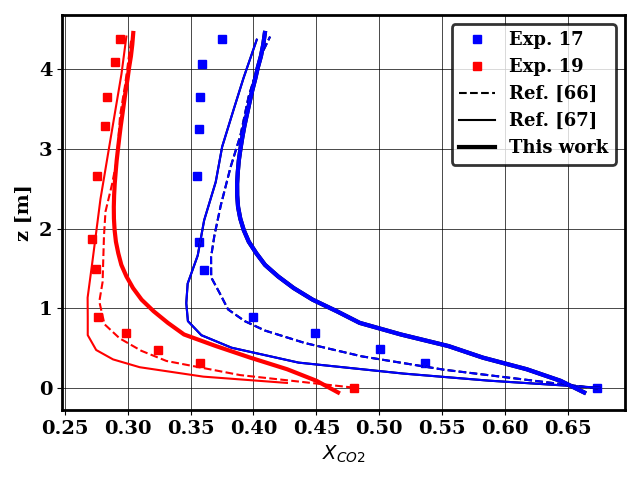}
    \caption{\revonetwo{Height conditional average of gas holdup (left) and \ce{CO2} mole fraction in the gas phase (right) for experiments (\mysquare{black}{black}) \cite{deckwer1978comprehensive}, Ref.~\cite{hissanaga2020mass} (\mydashedline{black}), Ref.~\cite{ngu2022spatio} (\myline{black}) and our results (\mythickline{black}), for Exp.~17 (blue) and Exp.~19 (red).}}
    \label{fig:singleLine}
\end{figure}

The experimental measurements provide the average local gas holdup and \ce{CO2} gas concentration at different heights in the reactor. In the simulations, these quantities are reconstructed by performing a conditional average of gas volume fraction and \ce{CO2} mole fraction conditioned on the height coordinate (z). The conditional average allows to eliminate radial variability at a given cross-section of the reactor. Figure~\ref{fig:singleLine} shows the comparison between simulations that were run with a global breakup model \cite{laakkonen2007modelling} and the experiments. For both the gas holdup and the \ce{CO2} gas mole fraction, the base simulation reasonably captures the experimental observations. \revone{Note that the conditional averaging operation smooths out the gradients of \ce{CO2} mole fraction and gas hold up which explains the discrepancy at the inlet.}

However significant differences can be observed especially for Exp.~19 for gas holdup and for Exp.~17 for \ce{CO2} gas concentration, especially near the top of the reactor. Other numerical investigations \cite{hissanaga2020mass,ngu2022spatio} where a one-dimensional spatio-temporal approximation of the phase conservation equations was used, also deviate from experimental observations in a similar way. Several investigations have observed that refining modeling strategies may not always close the gap between experimental observations and numerical simulations. For example, Huang et al.~\cite{huang2018assessment} shows that a population balance modeling approach for bubble size may not always be beneficial as compared to constant bubble size assumptions.  
It should also be noted that the conversion of \ce{CO2} to carbonates and bicarbonates could play an important role but gas-to-liquid \ce{CO2} mass transfer rates are typically much faster than these reaction rates, as described by Yue et al. \cite{yue2007hydrodynamics}, which would mean that gas-to-liquid mass transfer is the dominant physics in these experiments.

\subsection{Source of discrepancy between numerical simulation and the experiments}
\label{sec:disc}

\begin{figure}[h]
    \centering
    \includegraphics[width=0.4\textwidth]{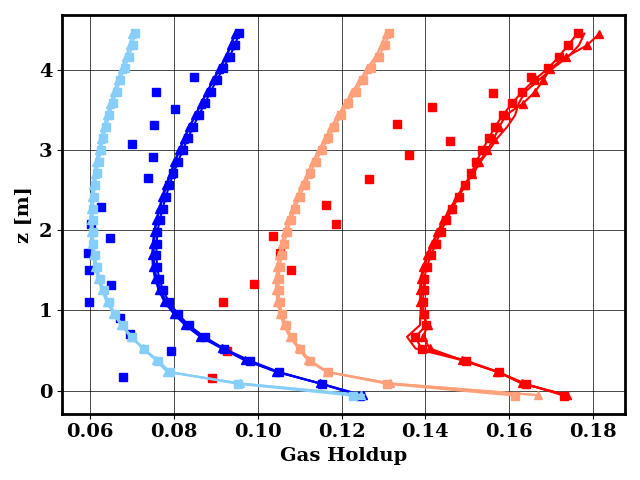}
    \includegraphics[width=0.4\textwidth]{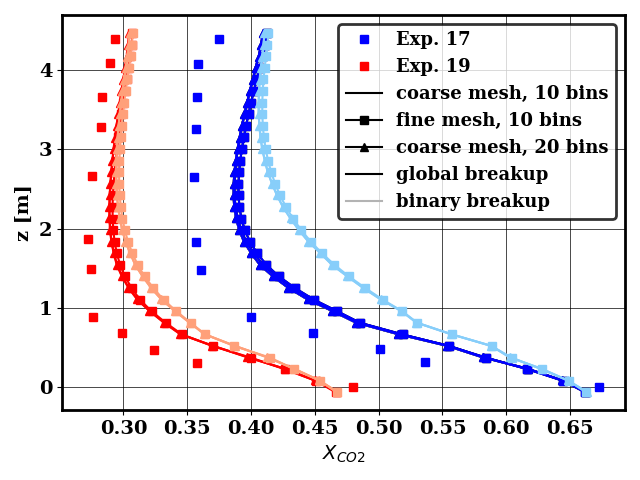}
    \caption{\revonetwo{Height conditional average of gas holdup (left) and \ce{CO2} mole fraction in the gas phase (right) for experiments (\mysquare{black}{black}) \cite{deckwer1978comprehensive}, coarse grid (\mythickline{black}), fine grid (\mythickbarredsquare{black}{black}), coarse grid with fine bubble size discretization  (\mythickbarredtriangle{black}{black}), for Exp. 17 (blue) and Exp. 19 (red). Dark lines: global breakup model \cite{laakkonen2007modelling}. Light lines: binary breakup model \cite{lehr2002bubble}.}}
    \label{fig:numerical}
\end{figure}

Two reasons might explain the deviation between numerical simulations and experiments. First numerical errors may originate from the discretization of the domain and must be ruled out before considering model modifications \cite{maluta2021effect}. A fine grid simulation was conducted using 176,400 cells (400 cells in the vertical direction, 28 in the azimuthal direction, and approximately 16 in the radial direction) compared to our baseline case with 64,800 cells. Additionally, a fine phase-space simulation was run using 20 bins for bubble-size discretization compared to 10 bins in the baseline case. In the fine grid simulation, the timestep is also based on the Courant number and is $1~\rm{ms}$ on average. As shown in Fig.~\ref{fig:numerical}, neither simulation deviated sufficiently from the base case to explain the discrepancy with experimental observations. Therefore, model errors are the likely cause of the discrepancy.

The same suite of runs (coarse and fine grids) was conducted with a binary breakup model \cite{lehr2002bubble} instead of a global breakup model \cite{laakkonen2007modelling}. This test serves here to understand how much variability can be expected due to a bubble size dynamics modeling error (similar to the approach of \citet{mueller2018model}). It can be seen in Fig.~\ref{fig:numerical} that, once again, the numerical errors are unlikely to explain the discrepancy between numerical simulations and experiments. However, significant variability may be observed due to the modeling assumption adopted, and this variability may indeed close the gap to experimental observations.

\section{Model calibration}
\label{sec:model_cal}

In view of the results shown above, the modeling approach may explain the discrepancy with the experimental data. Therefore, we used the experiments of \citet{deckwer1978comprehensive} to calibrate the breakup and coalescence models, in line with recommendations from previous research~\cite{laakkonen2006validation,laakkonen2007modelling, singh2009population}. The differences presented in Figure~\ref{fig:numerical} suggest that adjusting a set of parameters might be enough to resolve part of the discrepancy between simulation and experiments. The magnitude of the breakage rate, coalescence rate, and surface tension are likely candidates for these corrections. In other terms, it would be a reasonable model correction to adjust the selected parameters if the objective is to accurately capture mass transfer and gas holdup. In particular, it is possible that adjusting other parameters could resolve the same discrepancy. What is unique to the present study is that 1) the forward simulations are expensive (each forward run requires 360 CPU hours) and many runs were necessary to rule out numerical errors \cite{maluta2021effect} and, 2) there appears to be irreducible discrepancies between some of the experimental data and numerical simulations. Therefore, the inclusion of experimental data in the calibration procedure requires taking into account those irreducible errors. In this section, the solutions to both issues are provided and exercised.\footnote{The calibration methods and the model surrogates described in this section are available in a companion repository (\href{https://github.com/NREL/BioReactorDesign/tree/v0.0.16/}{\revonerone{https://github.com/NREL/BioReactorDesign/tree/v0.0.16/ \cite{birdswr}}}).}

\subsection{Calibration problem}

Given the high computational expense required for evaluating the forward model, the calibration procedure should not involve too many parameters. The higher the dimension of the parameter space, the larger the number of forward model evaluations needed to explore the parameter space. Note that even if a surrogate model is used for the Bayesian inference, it would still need to be trained using data that span the entire parameter space \cite{hassanaly2023pinn1}. Even when using faster forward models, other investigations have reduced the high-dimensional parameter space by assuming no joint effect of subsets of the entire parameter set \cite{laakkonen2007modelling}.

In this work, we calibrated 3 parameters: an efficiency factor for the coalescence kernel $C_{\rm eff}$, an efficiency factor for the breakup kernel $B_{\rm eff}$, and the surface tension $\Omega$. The base coalescence model is that of \citet{lehr2002bubble}. For the breakup model, two separate corrections are derived for the global breakup model of \citet{laakkonen2007modelling} and the binary breakup model of \citet{lehr2002bubble}. 

The scaled coalescence and breakup kernels are expressed as

\begin{equation}
    C_{\rm{corr}, v', v-v'} = C_{\rm eff} C_{v', v-v'}
\end{equation}

and 

\begin{equation}
    \revextrarone{B_{\rm{corr}, v} = B_{\rm eff} \beta_{v,v'} B_{v'}},
\end{equation}

where $C_{\rm{corr}, v', v-v'}$ and $B_{\rm{corr}, v}$ are the corrected coalescence and breakup kernels and $C_{\rm eff}$ and $B_{\rm eff}$ are the unitless efficiency factors that scale the base models. 

The choice of efficiency factors to calibrate the coalescence and breakup kernel is motivated by three reasons: 1) it allows to modulate the amplitude of the breakup and coalescence kernel by only manipulating one parameter which allows using a reduced number of forward evaluations; 2) the efficiency factor can be defined for any kernel functional which allows comparing different kernels and by how much each one must be corrected; 3) the efficiency factors are easily interpretable since their effect is linear on the frequency of breakup or coalescence.

Finally, the choice of calibrating surface tension $\Omega$ is motivated by the fact that liquid surface tension can be modified by the presence of surfactants in fermentation media and thus affect BSD and the mass transfer. 

\begin{figure}[h]
    \centering
    \includegraphics[width=0.4\textwidth]{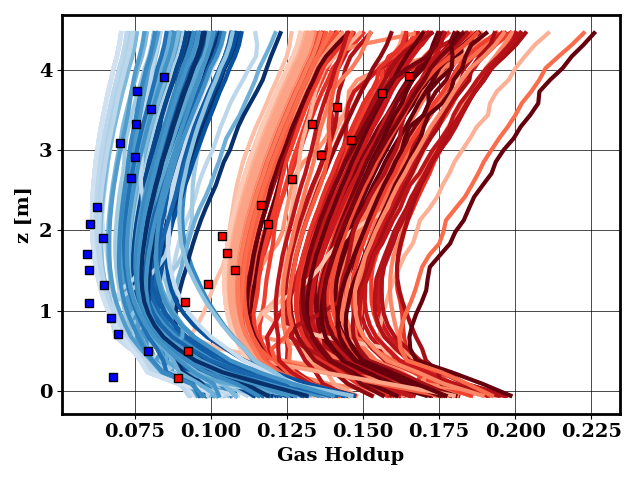}
    \includegraphics[width=0.4\textwidth]{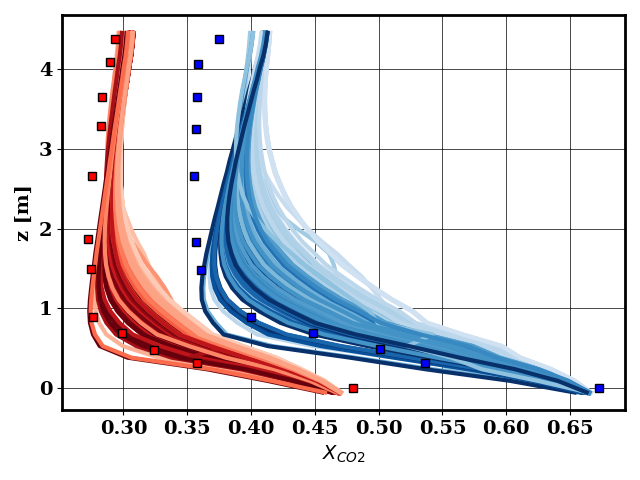}
    \includegraphics[width=0.4\textwidth]{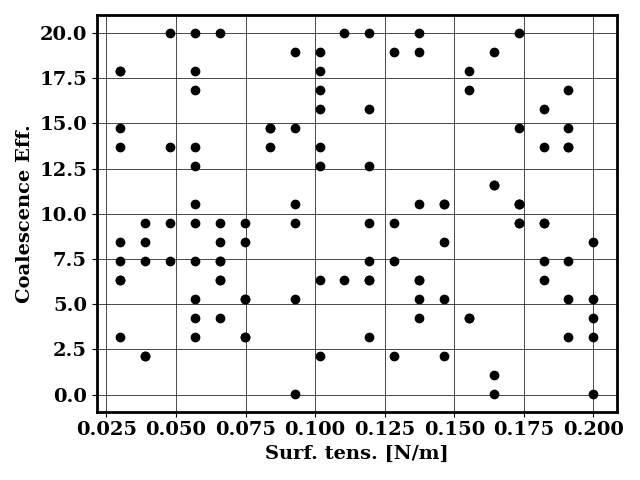}
    \includegraphics[width=0.4\textwidth]{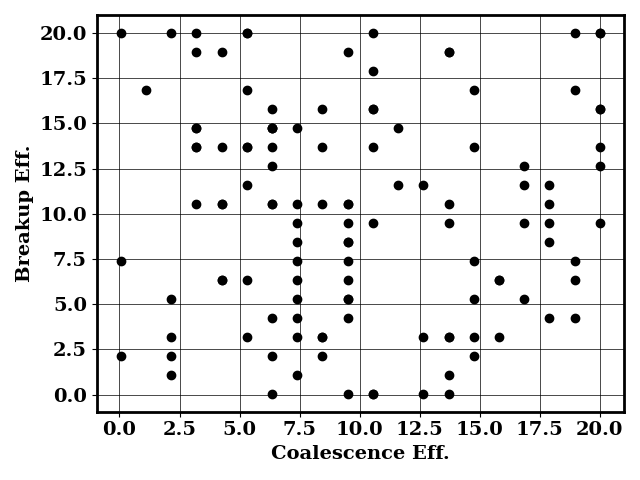}
    
    \caption{Top: Height conditional average of gas holdup (left) and \ce{CO2} mole fraction in the gas phase (right) for experiments (\mysquare{black}{black}) \cite{deckwer1978comprehensive}, and simulated results using a global breakup model \cite{laakkonen2007modelling} (\mythickline{black}) for Exp.~17 (blue) and Exp.~19 (red). Darker colors denote the higher value of breakup efficiency.
    Bottom: scatter plot of the distribution of the 120 numerical simulations in the model parameter space for the global breakup model.}
    \label{fig:shade}
\end{figure}

To decide what efficiency factors and surface tension values are most appropriate to explain the discrepancy between simulations and experimental observations, multiple simulations are run with different parameter values. A total of 120 numerical simulations were run with a coalescence/break-up efficiency factor varying in the range $[0.05, 20]$, and tap water surface tension varying in the range $[0.03, 0.11] \rm{N/m}$. \revtwo{In practice, the number of numerical simulations used to build the surrogate depends on the input space dimension and the smoothness of the function to approximate. Here, 120 simulations are equivalent to 5 discretization points per input dimension.} The results are shown in Fig.~\ref{fig:shade} for the global breakup and are colored by the breakup efficiency factor value, indicating that the computed gas holdup and \ce{CO2} mole fraction axial profiles are sensitive to the model parameters varied.

As suggested by the results of Sec.~\ref{sec:disc} the model parameter variation induces significant variability that could resolve part of the discrepancy to experimental observations. However, an optimal breakup efficiency factor is difficult to identify from Fig.~\ref{fig:shade}. On the one hand, high values of the breakup efficiency factor improve the \ce{CO2} mole fraction prediction, while on the other hand, low values of breakup efficiency promote a better match of gas holdup with experiments. In addition, there remain irreducible errors that cannot be mitigated through the variation of the parameters considered (see for instance gas \ce{CO2} concentration near the top of the reactor for Exp.~17). The same set of simulations resulted in similar conclusions when using the binary breakup kernel and is not described further for brevity.

\subsection{Bayesian inference approach}

To calibrate the chosen model parameters, while accounting for experimental uncertainty, a Bayesian calibration approach is used. In Bayesian calibration, the objective is to compute the posterior probability (or more commonly referred to as the posterior) of the parameters calibrated, denoted as $p_{\rm post}(\boldsymbol{p}|\boldsymbol{d})$, where $\boldsymbol{p}$ are the model parameters, and $\boldsymbol{d}$ are the experimental observations. The posterior $p_{\rm post}$ is obtained using the Bayes theorem
\begin{equation}
    p_{\rm post}(\boldsymbol{p}|\boldsymbol{d}) = \frac{p_{\rm prior}(\boldsymbol{p}) p_{\rm like} (\boldsymbol{d} | \boldsymbol{p})}{p_{\rm evi}(\boldsymbol{d})},
\end{equation}
where $p_{\rm prior}(\boldsymbol{p})$ is the prior probability that encodes prior knowledge about the parameters, and $p_{\rm like}(\boldsymbol{d} | \boldsymbol{p})$ is the likelihood function that characterizes how likely a parameter set $\boldsymbol{p}$ is to explain the observed data $\boldsymbol{d}$. The evidence $p_{\rm evi}(\boldsymbol{d})$ is shown for the sake of completeness but does not need to be defined for the Bayesian calibration procedure: it can be treated as a multiplicative factor independent of the parameters calibrated. Typically, the likelihood function is chosen to be a multivariate normal \cite{braman2013bayesian, hassanaly2023pinn2, bell2019bayesian} defined by 
\begin{equation}
\label{eq:like}
    \revextrarone{p_{\rm like}(\boldsymbol{d} | \boldsymbol{p})  = \frac{1}{(2\pi \sigma^2)^{N_{\rm d}/2}} \times \ \exp \left[ - \frac{1}{2 \sigma^2} \sum_{i=1}^{N_{\rm d}} ({d}_{\rm i} - {d}_{\rm i, pred}(\boldsymbol{p}))^2   \right]}, 
\end{equation}

where $N_{\rm d}$ is the number of observations, $\sigma$ is the uncertainty in the experimental observations (where uncertainty is assumed to be the same for all observations), \revextrarone{${d}_{\rm i, pred} (\boldsymbol{p})$ is the prediction of the $i^{th}$ observations that would be obtained if parameter set $\boldsymbol{p}$ was chosen, and  ${d}_{\rm i}$ is the $i^{th}$ observations.}

The advantage of the Bayesian calibration is that the experimental uncertainty and the remaining model uncertainty that cannot be explained with the parameters calibrated can be lumped into the $\sigma$ uncertainty value of the likelihood function \cite{hassanaly2023pinn2, smith2021Atikokan} (see Sec.~\ref{sec:comb}). Choosing a specific uncertainty for a specific experimental dataset allows to appropriately weight the experimental data, rather than averaging the error across all experimental data \cite{castellano2019using,mignard2006determination}.

\subsection{Surrogate modeling approach}

The main drawback of Bayesian calibration is its computational cost. The high cost of Bayesian calibration is due to evaluating the likelihood function. In the present case, it requires computing the prediction $\boldsymbol{d}_{\rm pred}(\boldsymbol{p})$ which would necessitate a forward model evaluation (here running a CFD simulation). To accelerate the Bayesian calibration, the function $\boldsymbol{d}_{\rm pred}(\boldsymbol{p})$ can, however, be approximated with a data-based model which was trained, in our case using the 120 simulations shown in Fig.~\ref{fig:shade}. Other uncertainty quantification approaches that require multiple forward model evaluations adopt similar strategies \cite{khalil2015uncertainty,raman2019emerging}.

\begin{figure*}[!h]
\centering
  \begin{subfigure}[b]{0.37\textwidth}
  \includegraphics[width=1.0\linewidth]{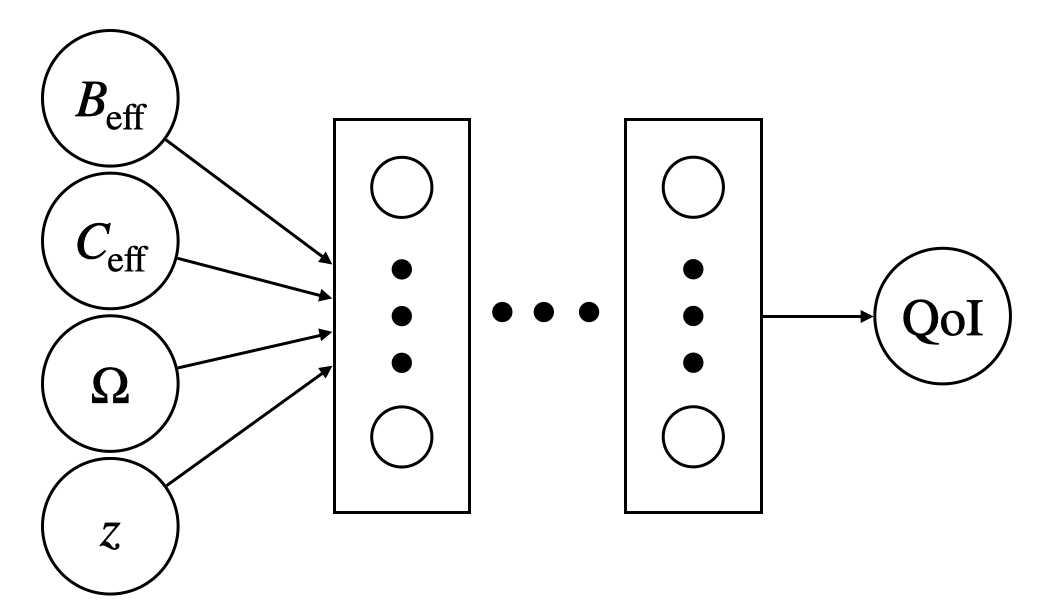} 
  \caption{Input output mapping of each surrogate model. Each surrogate takes 4 input and generates only one output denoted by QoI.}
  \label{fig:nnstruct}
  \end{subfigure}
  \hspace{3em}
  \begin{subtable}[b]{0.48\textwidth}
  \centering
    \begin{tabular}{l|l}
         Name & Value \\
         \hline
         Hidden layer size & \{ 10, 20, 10, 5\}  \\
         Activation & tanh \\
         Batch size & 32 \\
         Learning rate & $1\mathrm{e}{-03}$ \\
         Epochs & 1000         
    \end{tabular}
    \caption{Hyperparameters used for the neural net surrogates. The hidden layer size denotes the number of neurons in the 4 hidden layers used.}
    \label{tab:hyper}
  \end{subtable}
\caption{Architecture of the neural network surrogates.}
\label{fig:nn_arch}
\end{figure*}

Here, the surrogate models are constructed using a fully connected neural network schematically illustrated in Fig.~\ref{fig:nnstruct}. The neural net contains four inputs: the vertical direction $z$ and the three calibrated model parameters ($C_{\rm eff}$, $B_{\rm eff}$ and $\Omega$). The output of the neural net is one dimensional and consists of either the gas-holdup or the gas molar fraction of \ce{CO2}. In total, eight different surrogate models were trained, a different one for each measured quantity (gas hold up or \ce{CO2} mole fraction), each experiment (Exp.~17 or Exp.~19), and each breakup kernel (binary or global). For each surrogate model, 32 conditional average points are drawn for each simulation which amounts to 3840 data points per surrogate. One percent of the data is reserved for validation. For simplification, all surrogate models used the same hyperparameters that are summarized in Tab.~\ref{tab:hyper}. \revtwo{Given the small amount of training data, the number of trainable parameters needs to be small enough to prevent overfitting. Here, the number of hidden layers was fixed to 4 and the number of neurons in each layer was manually reduced until the validation loss reached levels similar to the training loss (\ref{app:convergenceNN}).} The neural nets were trained with a classical mean-square error loss. Further details on the effect of the epoch number are provided in \ref{app:convergenceNN}.

The surrogate model can be used to compute the likelihood function $p_{\rm like} (\boldsymbol{d} | \boldsymbol{p})$. To perform Bayesian calibration, the prior distribution of the parameters must also be chosen. In order to use the surrogate model, the support of the priors cannot be larger than the input feature intervals over which the surrogate was trained. The priors for the coalescence and the breakup efficiency factor are chosen to be uniform distributions $\mathcal{U}(0.05, 20)$, whose support is the same as the interval over which the surrogate was trained. In general, the magnitude of the breakup kernels adopted does not vary by a factor greater than 20 \cite{azizi2011turbulently}. The priors adopted reflect that scaling the coalescence and breakup kernel by a factor greater than 20 is unreasonable. The prior for the surface tension is chosen to be $\mathcal{U}(0.065, 0.075)$ N/m. Given available estimates of the surface tension of tap water \cite{stagonas2011surface,o1972purity}, the support of the surface tension prior was made tighter than the interval over which the surrogate was trained.

\subsection{Combining multiple observations}
\label{sec:comb}

\begin{figure}[h]
    \centering
    \includegraphics[width=0.4\textwidth]{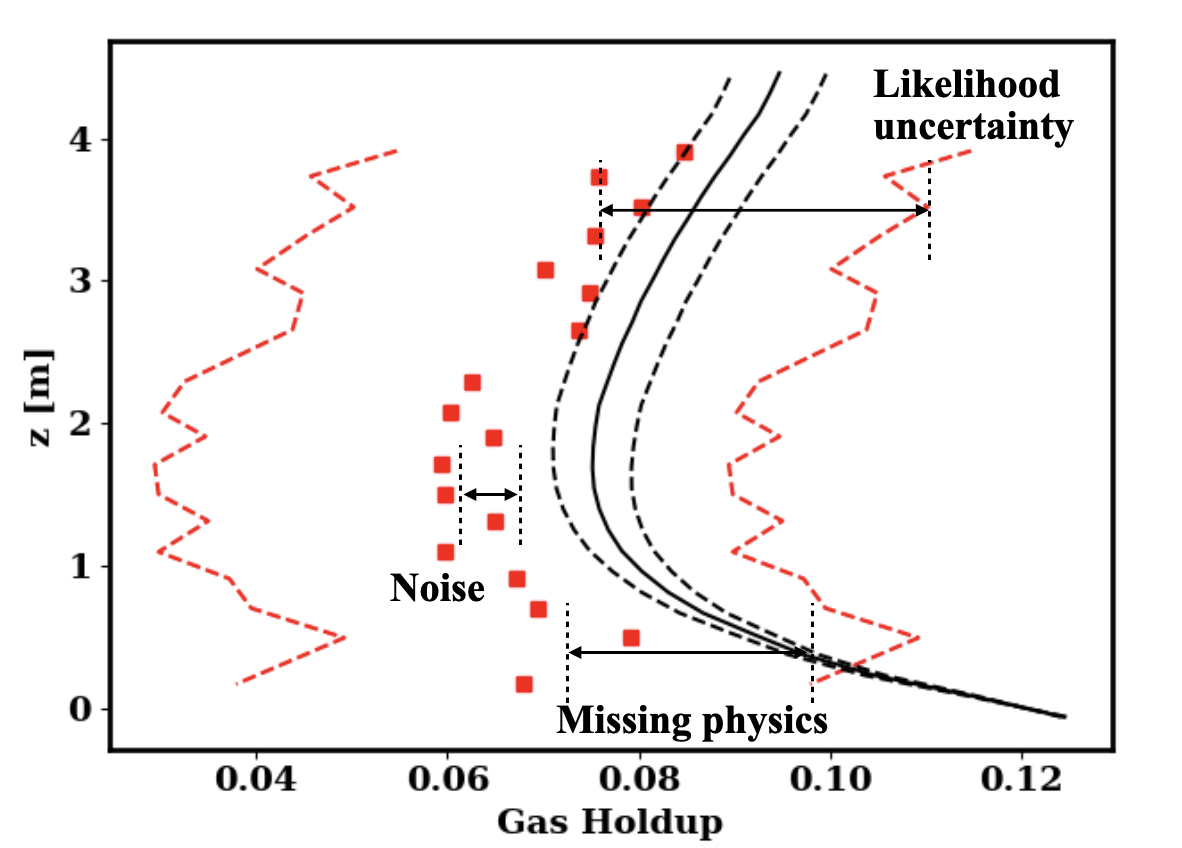}
    \caption{Illustration of the missing physics issue. If experimental noise and model parameter uncertainty (black dashed lines) are not necessary to explain discrepancies between simulation and experiments, a missing physics component is added to the likelihood uncertainty (red dashed lines).}
    \label{fig:missPhy}
\end{figure}

As mentioned in Sec.~\ref{sec:motivation}, it is common practice to calibrate bubble size dynamics models against the simulation of experimental observations, by uniformly weighting the errors to experimental data \cite{castellano2019using,mignard2006determination,solsvik2014population}. This approach implicitly assumes that all experiments are equally informative to calibrating model parameters. However, different experiments may be subject to different levels of noise, especially when different variables are measured. \revtwo{In practice, noise may be due to statistical uncertainty \cite{oliver2014estimating} or to non-controllable variability in macroscopic operating conditions (for example, volume flow rate or mean temperature), that affect different measurement types in distinct ways.} For instance, Fig.~\ref{fig:numerical} suggests that the noise-to-signal ratio is larger for the gas holdup compared to \ce{CO2} mole fraction. Furthermore, despite pushing the parameters to calibrate to their limit, there remain irreducible discrepancies between some experiments and simulation results (Fig.~\ref{fig:shade}). In these cases, to match such experimental observations, the models need to be equipped with additional physics that are missing. The missing physics could be because of inherent model deficiencies, or because the experiments were affected by unknown or undocumented variables. Either way, the missing physics makes the experimental observation less informative about the parameters being calibrated. Figure~\ref{fig:missPhy} schematically illustrates the missing physics issue. To account for the missing physics, the likelihood uncertainty $\sigma$ (Eq.~\ref{eq:like}) can be adjusted to reflect the fact some experiments are less informative for the parameter calibration. In Fig.~\ref{fig:missPhy}, the likelihood uncertainty is plotted as an experimental uncertainty (red dashed lined) that contains the noise and the missing physics. Hereafter, the likelihood uncertainty is assumed uniform for each one of the four experimental datasets: a single likelihood uncertainty is used for each measurement type ($X_{\ce{CO2}}$, gas holdup) and each experimental campaign (Exp.~17 and Exp.~19).

To set the likelihood uncertainty so as to reflect both the experimental noise and the missing physics, two different methods are used. First, the Bayesian calibration can be done for each one of the four experimental datasets while calibrating the likelihood uncertainty along with the physical parameters (coalescence efficiency, breakup efficiency, surface tension). As a second step, the mean likelihood uncertainty of each one of the four experimental datasets is gathered and used to perform the final calibration that combines all experimental data. This procedure is summarized in Algo.~\ref{algo:calUnc} where $\mathbb{E}_{p_{\rm post}}(\sigma)$ denotes the average of the likelihood $\sigma$ averaged over the posterior probability sampled.
 
\begin{algorithm}
\caption{Likelihood uncertainty calibration.}
\label{algo:calUnc}
\begin{algorithmic}[1]
\For{$\text{Exp}_{\rm i}$ $\in$ \{ Exp.~17 Gas Holdup, Exp.~17 $X_{\ce{CO2}}$, Exp.~19 Gas Holdup, Exp.~19 $X_{\ce{CO2}}$ \}}
    \State Calibrate $B_{\rm eff}$, $C_{\rm eff}$, $\Omega$ and $\sigma$
    \State Compute $\sigma_i = \mathbb{E}_{p_{\rm post, i}}(\sigma)$
\EndFor
\end{algorithmic}
\end{algorithm}

Second, the likelihood uncertainty can be treated as a hyperparameter that needs to be optimized. Qualitatively, the uncertainty band obtained by calibrating the model parameters (black dashed lines in Fig.~\ref{fig:missPhy}) must be contained in the likelihood uncertainty assigned to the experiments (the red dashed lines in Fig.~\ref{fig:missPhy}). The definition of uncertainty bands is arbitrary and is chosen to be the $95\%$ confidence interval for the experiments, and the $95\%$ confidence interval for the model predictions. Formally, the likelihood uncertainty $\sigma_i$ is chosen by solving the following optimization problem for each experiment $i$

\begin{equation}
\label{eq:sigOpt}
\begin{aligned}
\min_{\sigma_i}  \\
\textrm{s.t.} \quad & V_{i,j} + 2 \sigma_i > P_{i, j, 97.5} (\sigma_i),~ \forall j\\
  & V_{i,j} - 2 \sigma_i < P_{i, j, 2.5} (\sigma_i), ~ \forall j,    \\
\end{aligned}
\end{equation}

where $V_{i,j}$ is the $j^{th}$ measurement of experimental dataset $i$, $P_{i,j, 97.5}$ (resp. $P_{i,j,2.5}$) is the 97.5$^{th}$ (resp. 2.5$^{th}$) percentile of the model predictions for the $j^{th}$ measurement of experiment $i$ obtained by sampling the posterior of  $B_{\rm eff}$, $C_{\rm eff}$ and $\Omega$.

The optimal value of the likelihood uncertainty can be iteratively searched, where each iteration consists in choosing a value of $\sigma_i$ and calibrating the model parameters  ($B_{\rm eff}$, $C_{\rm eff}$, $\Omega$) using $\sigma_i$ as likelihood uncertainty.  Since a surrogate model is used in place of the physics model, the calibration process is sufficiently cheap to allow performing multiple calibrations. The procedure is summarized in Algo.~\ref{algo:optUnc}. \revone{A validation procedure in a simple illustrative case is provided in ~\ref{app:val}.}

\begin{algorithm}
\caption{Likelihood uncertainty optimization.}
\label{algo:optUnc}
\begin{algorithmic}[1]
\For{$\text{Exp}_{\rm i}$ $\in$ \{ Exp.~17 Gas Holdup, Exp.~17 $X_{\ce{CO2}}$, Exp.~19 Gas Holdup, Exp.~19 $X_{\ce{CO2}}$ \}}
    \State Solve Eq.\ref{eq:sigOpt} via a bisection method 
    \State Store the solution $\sigma_i$ for $\text{Exp}_{\rm i}$.
\EndFor
\end{algorithmic}
\end{algorithm}

The likelihood uncertainties obtained with both methods are shown in Tab.~\ref{tab:likePhy} for the Binary breakup and the global breakup model. Overall, the uncertainties predicted with both methods are in agreement with each other, which suggests that calibrating likelihood uncertainties is equivalent to solving the optimization problem shown in Eq.~\ref{eq:sigOpt}. Importantly, calibrating the likelihood uncertainties only requires one calibration procedure per experiment, while 10 were used for the hyperparameter search.

The likelihood uncertainties reflect the amount of noise in the experimental data as well as the amount of physics missing to explain the experiments with the models. By comparing the likelihood uncertainties across modeling assumptions (global breakup or binary breakup assumption), one can compare the amount of physics missing in each model, irrespective of the calibrated model parameters. This is possible here since the breakup model is the only modeling choice varied, while the coalescence model is the same. Other Bayesian model ranking strategies have been proposed \cite{braman2013bayesian} by accounting for the amount of parameter tuning needed for each model. A simpler model ranking method is used here since both models need to be tuned by the same amount (see Sec.~\ref{sec:res}).
Here, it appears that the global breakup model is about two times more accurate than the binary breakup model for $X_{\ce{CO2}}$ and is only 10-20\% less accurate than the binary breakup model for gas holdup.

\begin{table}[h]
    \vspace{2em}
    \begin{subtable}[h]{0.49\textwidth}
            \centering
        \scriptsize
        \setcellgapes{5pt}
        \makegapedcells
        \vspace*{-5mm}
        
        \begin{tabular}{ |c|c|c|c|c| } 
        \hline
        &  \multicolumn{2}{|c|}{\textbf{Exp. 17}} & \multicolumn{2}{|c|}{\textbf{Exp. 19}} \\
        \hline
         & $X_{CO_2}$ & Gas Holdup & $X_{CO_2}$ & Gas Holdup \\
        \hline
        Binary Breakup & 0.058 & 0.009 & 0.03 & 0.016  \\
        \hline
        Global Breakup & 0.034 & 0.008 & 0.015 & 0.02  \\
        \hline
        \end{tabular}
       \caption{Likelihood uncertainties obtained using Algo.~\ref{algo:calUnc}.}
       \label{tab:likeCal}
     \end{subtable}
     \hfill
    \begin{subtable}[h]{0.49\textwidth}
        \centering
        \scriptsize
        \setcellgapes{5pt}
        \makegapedcells
        \vspace*{-5mm}
        
        \begin{tabular}{ |c|c|c|c|c| } 
        \hline
        &  \multicolumn{2}{|c|}{\textbf{Exp. 17}} & \multicolumn{2}{|c|}{\textbf{Exp. 19}} \\
        \hline
         & $X_{CO_2}$ & Gas Holdup & $X_{CO_2}$ & Gas Holdup \\
        \hline
        Binary Breakup & 0.066 & 0.009 & 0.039 & 0.024  \\
        \hline
        Global Breakup & 0.025 & 0.010 & 0.015 & 0.027  \\
        \hline
        \end{tabular}
       \caption{Likelihood uncertainties obtained using Algo.~\ref{algo:optUnc}.}
       \label{tab:likeOpt}
    \end{subtable}

     \caption{Optimal likelihood uncertainty for each experimental dataset and each breakup model.}
     \label{tab:likePhy}
\end{table}

Given that interphase mass transfer is a critical parameter for \ce{CO2} gas fermenters, accurate predictions of the gas mole fraction $X_{\ce{CO2}}$ are needed. The calibrated and optimized likelihood uncertainties suggest that irrespective of efficiency parameter calibration, a global breakup model \cite{laakkonen2007modelling} is more appropriate to capture the experimental interphase mass transfer observed compared to the binary breakup model \cite{lehr2002bubble}.

\subsection{Results}
\label{sec:res}

\subsubsection{Parameters' posteriors}
\label{sec:postpar}
\begin{figure}[h]
    \centering
    \includegraphics[width=0.49\textwidth]{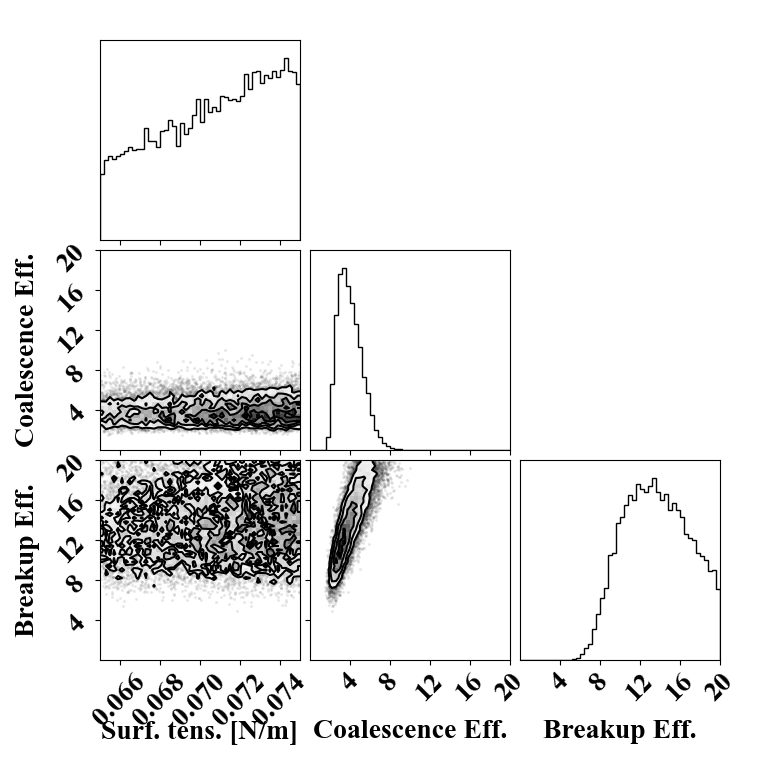}
    \includegraphics[width=0.49\textwidth]{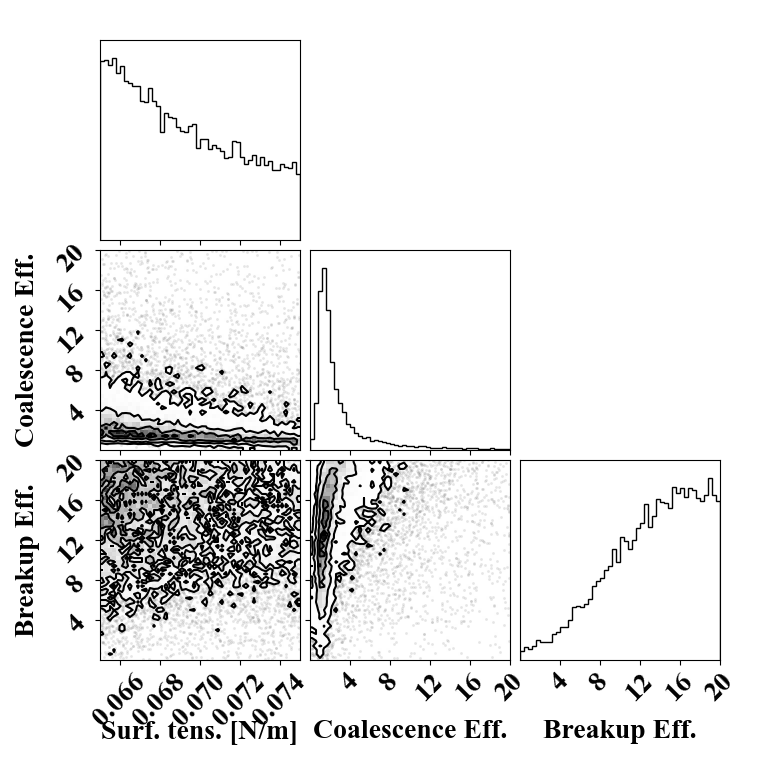}
    \caption{Corner plots of the posterior probability obtained for the global breakup model \cite{laakkonen2007modelling} (left) and the binary breakup model \cite{lehr2002bubble} (right).}
    \label{fig:corner}
\end{figure}

The Bayesian calibration is conducted using Hamiltonian Monte-Carlo (HMC) \cite{hoffman2014no, numpyro} which is made possible due to the auto-differentiation capability of the neural network surrogate. The likelihood uncertainties obtained with the calibration (Algo.~\ref{algo:calUnc}) and the optimization method  (Algo.~\ref{algo:optUnc}) are used to compute the likelihood function that combines all the experiments. One calibration is done with the global breakup model and one with the binary breakup model. Figure~\ref{fig:corner} shows the posterior probability density function (PDF) of the calibrated parameters obtained when combining all the experiments and using the likelihood uncertainties computed via the optimization method  (Algo.~\ref{algo:optUnc}). The posteriors obtained when using the likelihood uncertainties computed via the calibration method  (Algo.~\ref{algo:calUnc}) are similar and are not shown here for the sake of concision.

First, it appears that any value of surface tension within the interval chosen is reasonable to explain the discrepancy with experimental data. Therefore, the experimental data is not sufficient to confidently decide whether the surface tension of tap water was affected by surfactants. \revtwo{This result is confirmed by a separate sensitivity analysis which shows that the predicted gas holdup and gaseous \ce{CO2} mole fraction are more sensitive to breakup and coalescence efficiency than surface tension (\ref{app:sens}).} However, only narrow regions of the coalescence and breakup efficiency domain of definition are likely. The coalescence efficiency is found most likely between 2 and 4, while the breakup efficiency is found most likely between values 10 and 15. Remarkably, both breakup models (global and binary) land on similar conclusions regarding the posterior probability. The posterior exhibits a correlation between coalescence and breakup efficiencies which can be understood by the fact that higher breakup rates may be compensated by higher coalescence rates. The agreement of the calibrated efficiency factors suggests that the model correction are a) either needed for modeling the bubble dynamics of gaseous \ce{CO2}; or b) needed to model the specific experiment simulated here. A definitive conclusion would require performing the same calibration exercise with other experimental measurements and is left for future work. \revtwo{Finally, the most likely parameter values are not located at the prior bounds (except for breakup efficiency of the binary breakup case). Therefore, further extending the support of the priors is unlikely to yield a significant change in the calibration results.}

\begin{figure}[h]
    \centering
    \includegraphics[width=0.49\textwidth]{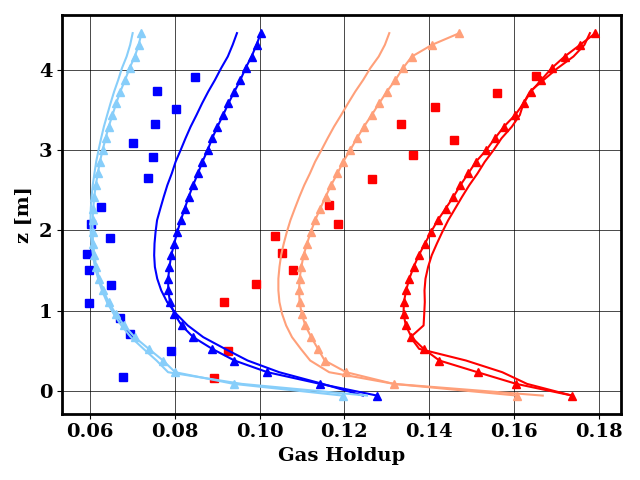}
    \includegraphics[width=0.49\textwidth]{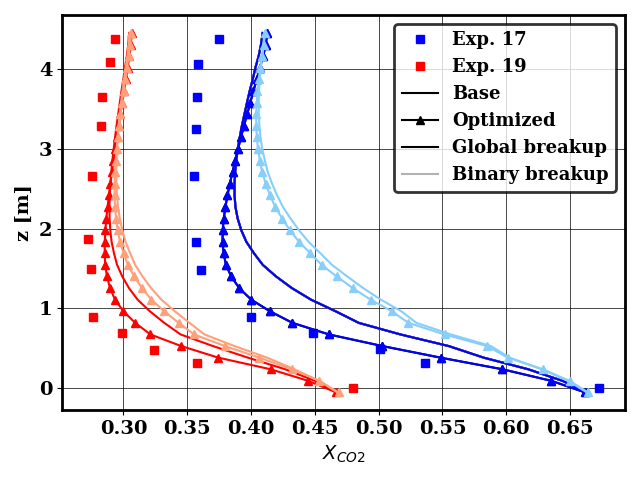}
    \caption{\revonetwo{Height conditional average of gas holdup (left) and \ce{CO2} mole fraction in the gas phase (right) for experiments (\mysquare{black}{black}) \cite{deckwer1978comprehensive}, the base model (\mythickline{black}), the calibrated model (\mythickbarredtriangle{black}{black}) for Exp. 17 (blue) and Exp. 19 (red). Dark lines: global breakup model \cite{laakkonen2007modelling}. Light lines: binary breakup model \cite{lehr2002bubble}.}}
    \label{fig:predopt}
\end{figure}

A last set of simulations, termed \textit{a posteriori} runs, was performed to verify the predictions of the optimized models. A binary breakup model was used with coalescence efficiency factor $C_{\rm eff } = 1.83$ and breakup efficiency factor $B_{\rm eff } = 13.565$ and a global breakup model was used with $C_{\rm eff } = 4.695$ and $B_{\rm eff } = 13.83$. The values chosen correspond to the posterior means of $B_{\rm eff }$ and $C_{\rm eff }$. Figure~\ref{fig:predopt} shows the predicted results of the optimized models against the base models using the baseline computational grid and bubble size discretizations (Sec.~\ref{sec:disc}). In the case of the binary breakup model, the optimized model mostly affected the gas holdup prediction, especially for Exp.~19. However, the gas mole fraction concentrations are left almost unchanged. This is a consequence of the high likelihood uncertainty for $X_{\ce{CO2}}$ \revone{(relative to the value of $X_{\ce{CO2}}$)} which leads to prioritize gas holdup in the model calibration. The opposite effect can be seen for the global breakup model where significant improvement can be observed for $X_{\ce{CO2}}$ while gas holdup is unchanged or even less accurate (especially for Exp.~17). Overall, the calibration using the neural network surrogate is consistent with the a posteriori runs, which confirms that the neural network surrogate was sufficiently accurate. \revone{The effect of the calibrated model can also be evaluated with the BSD and is shown in \ref{app:effectBSD}.} \revone{Finally, since the same target dataset is used to calibrate the binary breakup and the global breakup model, one could have expected that both calibrated models match. However, the missing physics uncertainty quantification procedure weighs each experimental dataset differently for each breakup model, which explains why the calibrated results differ.} 

\subsubsection{\revonerone{Test for general applicability}}

\begin{figure}[h]
    \centering
    \includegraphics[width=0.90\textwidth]{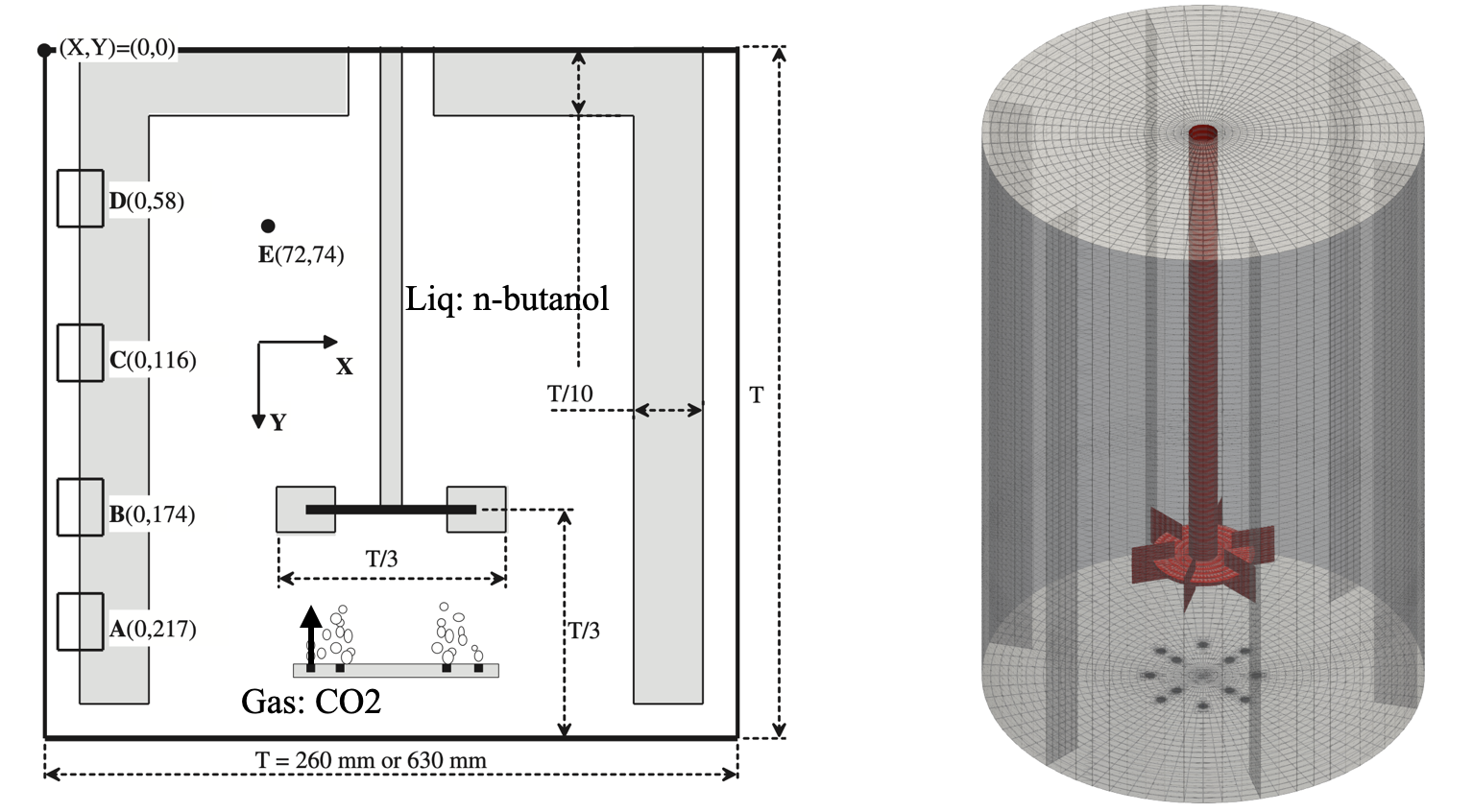}
    \caption{Left: schematic illustration of \revone{stirred-tank reactor} simulated adapted from \cite{laakkonen2007modelling}. Right: computational mesh used for simulating the \revone{stirred-tank reactor}.}
    \label{fig:lakkonen_ill}
\end{figure}

The parameter calibration procedure was performed over a single experimental investigation in order to capture gas holdup an interphase mass transfer only. In general, it is not possible to guarantee that the model parameters will accurately predict any other variable, or that the model parameters will be applicable to other reactor types. The results of Sec.~\ref{sec:postpar} suggest, however, that the appropriate efficiency factors to use are similar even for different breakup model assumptions, which points to a general applicability of the parameters identified. In this section, the applicability of the optimal correction factors \revtwo{for breakup and coalescence efficiency} is tested in a different case described in Laakkonen et al.~\cite{laakkonen2007modelling}. \revtwo{Surface tension is not adjusted because it is a model input that depend of the experiment conducted.} The configuration simulated is the 14L \revone{stirred-tank reactor} graphically illustrated in Fig.~\ref{fig:lakkonen_ill} (left with T=$260$ mm). The reactor is operated with pure gaseous \ce{CO2} and pure liquid n-butanol. The rotation speed of the impeller (shown in red in Fig.~\ref{fig:lakkonen_ill} right) is 700 rpm, and the mass flow rate of gaseous \ce{CO2} is set to 0.7 vvm. To simplify the meshing, the baffles, and the propeller blades were modeled
as zero-thickness walls. The inlet is constructed with 12 holes of diameter 1 mm. The mesh is refined near the inlet holes as shown in Fig.~\ref{fig:lakkonen_ill} (right). The smallest cell has an edge length of 0.1 mm and the largest has an edge length of 6.2 mm. In total, the mesh contains 2.2 million cells. Similar to Sec.~\ref{sec:numerics}, a no-slip boundary condition is used at the walls for the liquid phase, and a slip boundary condition is used at the wall for the gas phase.
The BSD is discretized using 14 bins equally spaced ranging from 0.1 mm to 4 mm. Similar to Sec.~\ref{sec:numerical}, the timestep was dynamically adjusted based on the maximal Courant number, and was $2.5~\rm{\mu s}$ on average. The $k$-$\varepsilon$ model was used (consistently with Sec.~\ref{sec:numerical}) and the inlet boundary conditions for $k$ and $\varepsilon$ were based on freestream turbulence.

In Ref.~\cite{laakkonen2007modelling}, the bubble size distribution is reported at 4 different windows in the reactor (also shown in Fig.~\ref{fig:lakkonen_ill} left). The n-butanol solvent was chosen to ensure minimal interphase mass transfer. Here, mass transfer is modeled and the Henry's constant is obtained from \cite{kvam2019solubility} where it is reported for an ethanol liquid phase (instead of n-butanol). As a separate analysis, not reported here, the effect of this assumption was tested by canceling the interphase mass transfer entirely which resulted in minimal impact on the quantities of interest (the bubble size distribution). Similar to Sec.~\ref{sec:numerical}, the model of \citet{lehr2002bubble} is used for coalescence, and the model of \citet{laakkonen2007modelling} is used for breakup. Two cases are simulated: case 1 uses a coalescence and breakup efficiency factor of $1$; case 2 uses a coalescence efficiency factor of $C_{\rm eff } = 4.695$ and $B_{\rm eff } = 13.83$, consistently with Sec.~\ref{sec:postpar}.

\begin{figure}[h]
    \centering
    \includegraphics[width=0.7\textwidth]{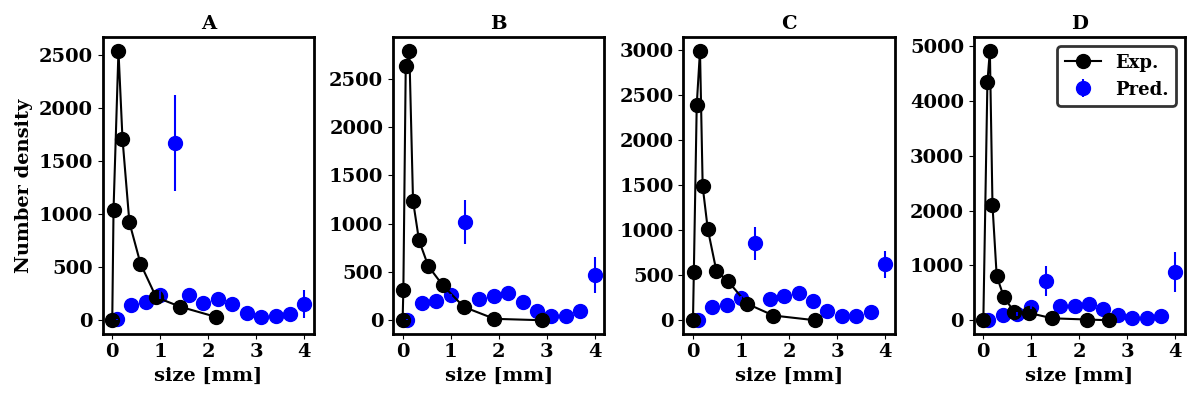}
    \includegraphics[width=0.7\textwidth]{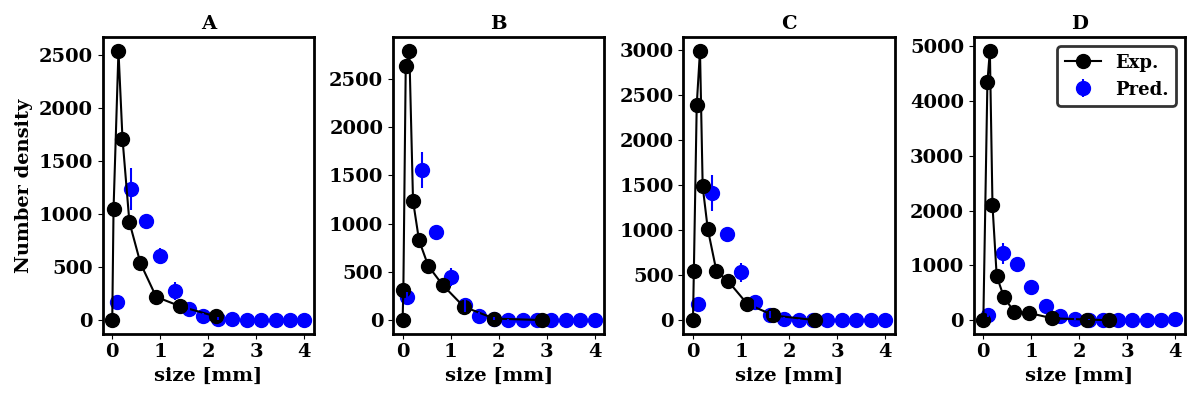}
    \caption{\revonetwo{Bubble size distribution at the 4 measurement windows from experimental measurements \cite{laakkonen2007modelling} (\mycircle{black}{black}) and from the present simulations (\mycircle{blue}{blue}). Error bars denote a one standard deviation variability over the spatial averaging window. Top: case 1. Bottom: case 2.}}
    \label{fig:bsd}
\end{figure}

Figure~\ref{fig:bsd} shows the bubble size distribution results for case 1 and case 2 at steady state. Although the efficiency factors were not calibrated to capture BSD, it is clear that they significantly improve the predicted BSD. In particular, Fig.~\ref{fig:bsd} shows that case 1 severely underestimates the bubble breakup. Note that in \citet{laakkonen2007modelling}, the breakage model was derived in tandem with a coalescence model to match the bubble size distribution. Here, since only the breakage model from \citet{laakkonen2007modelling} is used, the same output cannot be expected. For case 2, the accuracy of the BSD is significantly improved and the mismatch observed is on par with other analyses \cite{lehnigk2022open}.

    



\section{Conclusions}
\label{sec:conclusions}

In this work, an inverse modeling procedure was conducted to calibrate bubble size dynamics models against experimental measurements of a coflowing bubble column reactor.

Experimental noise and simulation bias were accounted for using a Bayesian inference approach. Given the common practice of calibrating bubble size dynamics model parameters, choosing an appropriate model functional form, irrespective of the parameter calibration is non-trivial. Here, it is shown that estimating the simulation bias allows one to discriminate between modeling approaches. In the present case, a global breakup modeling approach proved more appropriate for capturing interphase mass transfer, which is essential for gas fermentation applications. The simulation bias can be estimated either by including the likelihood uncertainty in the set of calibrated parameters, or by treating the likelihood uncertainty as a hyperparameter.

For computational tractability, the Bayesian inference needs to be paired with a surrogate model rather than a physics model. It was shown that a simple neural network was able to approximate the dependence of the model prediction with respect to the model parameters inferred.  \revtwo{The computational approach proposed is applicable to gas-liquid systems in general where parameters, subject to uncertainty, are used for closure modeling.}

The calibration results show that for both global breakup and binary breakup modeling approaches, the magnitude of the breakup kernel needs to be increased by about one order of magnitude in order to accurately capture gas holdup and interphase mass transfer. 

\revtwo{The specific parameter adjustments proposed in this work do not necessarily apply to all gas-liquid systems, \revonerone{but the tests in this work indicate} general applicability.}
\revtwo{In particular, calibrated parameters were shown to be beneficial to predict bubble size distribution in an unseen \revone{stirred-tank reactor} case}. Since both breakup modeling approaches led to similar conclusions, and are also appropriate for a test case not used to calibrate the parameters, it is reasonable to suspect a physical reason behind the identified efficiency factors. Either the breakup dynamics of \ce{CO2} gas are atypical, or an element of both experimental apparatuses promoted high breakage rates. A definitive conclusion would require exercising the calibration method described against a larger experimental dataset and additional model parameters. 


\section*{Acknowledgments}
The authors thank Julie Bessac for fruitful discussions.
This work was authored by the National Renewable Energy Laboratory (NREL), operated by Alliance for Sustainable Energy, LLC, for the U.S. Department of Energy (DOE) under Contract No. DE-AC36-08GO28308. This work was supported by the U.S. Department of Energy Office of Energy Efficiency and Renewable Energy Bioenergy Technologies Office (BETO). The research was performed using computational resources sponsored by the Department of Energy's Office of Energy Efficiency and Renewable Energy and located at the National Renewable Energy Laboratory. The views expressed in the article do not necessarily represent the views of the DOE or the U.S. Government. The U.S. Government retains and the publisher, by accepting the article for publication, acknowledges that the U.S. Government retains a nonexclusive, paid-up, irrevocable, worldwide license to publish or reproduce the published form of this work, or allow others to do so, for U.S. Government purposes.

\appendix
\label{sec:appendix}

\section{Convergence of the neural networks}
\label{app:convergenceNN}

\begin{figure}[h]
    \centering
    \includegraphics[width=0.4\textwidth]{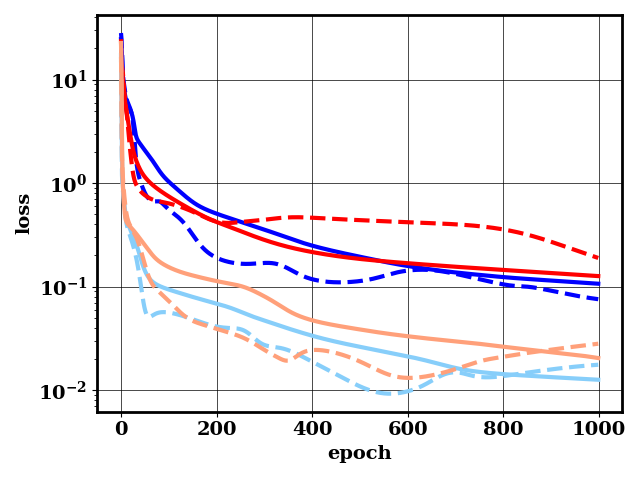}
    \includegraphics[width=0.4\textwidth]{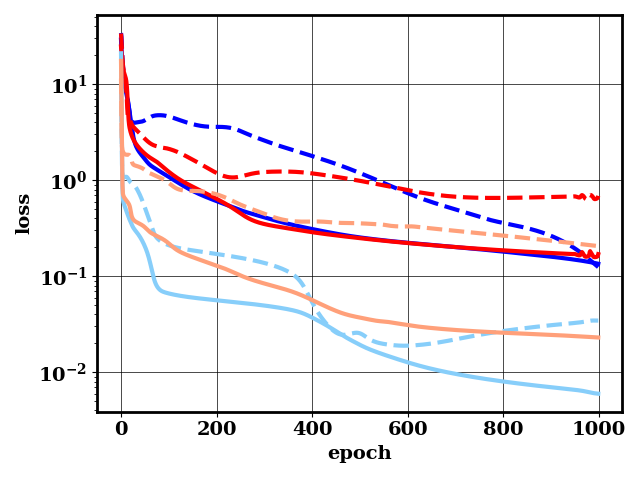}
    \caption{Training (\mythickline{black}) and testing (\mythickdashedline{black}) loss history of the neural network surrogates trained on the Exp.~17 (\mythickline{blue}) and Exp.~19 (\mythickline{red}), for the global breakup model \cite{laakkonen2007modelling} (left) and the binary breakup model \cite{lehr2002bubble} (right). Dark lines: gas holdup. Light lines: \ce{CO2} gas mole fraction.}
    \label{fig:conv}
\end{figure}

Figure~\ref{fig:conv} shows the training and validation loss history of the neural network surrogate of the breakup model and of the binary breakup model. Overall, it can be seen that the training and testing loss are nearly converged. Since the testing losses do not gradually increase with the epoch number, the surrogates are unlikely to overfit.

\section{Turbulence boundary conditions}
\label{app:bcke}

The turbulent boundary conditions for the turbulent kinetic energy $k$ and the turbulent energy dissipation rate $\varepsilon$ were artificially increased in the present work (Sec.~\ref{sec:comp_exp}, Sec.~\ref{sec:disc} and Sec.~\ref{sec:model_cal}) for stability reasons. The chosen turbulent inlet boundary conditions are reported in Tab.~\ref{tab:bc}, along with theoretical values for freestream turbulence, assuming 5\% turbulence intensity in the liquid phase and the gas phase. 

\begin{table}[h]
    \vspace{2em}
    \begin{subtable}[h]{0.49\textwidth}
            \centering
        \scriptsize
        \setcellgapes{5pt}
        \makegapedcells
        \vspace*{-5mm}
        
        \begin{tabular}{ |c|c|c|c|c| } 
        \hline
        &  \multicolumn{2}{|c|}{\textbf{Exp. 17}} & \multicolumn{2}{|c|}{\textbf{Exp. 19}} \\
        \hline
         & Gas & Liquid & Gas & Liquid \\
        \hline
        This work &  3.7e-5 & 0.01 &  3.7e-5 & 0.01 \\
        \hline
        Turbulence intensity 5\%  &  4.4e-6 & 8.3e-6 & 8.0e-6 & 8.3e-6  \\
        \hline
        \end{tabular}
       \caption{Turbulent kinetic energy boundary conditions in $m^2.s^{-2}$.}
       \label{tab:kbc}
     \end{subtable}
     \hfill
    \begin{subtable}[h]{0.49\textwidth}
        \centering
        \scriptsize
        \setcellgapes{5pt}
        \makegapedcells
        \vspace*{-5mm}
        
        \begin{tabular}{ |c|c|c|c|c| } 
        \hline
        &  \multicolumn{2}{|c|}{\textbf{Exp. 17}} & \multicolumn{2}{|c|}{\textbf{Exp. 19}} \\
        \hline
         & Gas & Liquid & Gas & Liquid \\
        \hline
        This work & 1.5e-4 & 1.5e-4 & 1.5e-4 & 1.5e-4  \\
        \hline
        Turbulence intensity 5\% & 1.44e-7 & 3.75e-7 & 3.6e-7 &3.75e-7  \\
        \hline
        \end{tabular}
       \caption{Turbulent kinetic energy dissipation rate boundary conditions in $m^2.s^{-3}$.}
       \label{tab:ebc}
    \end{subtable}

     \caption{Comparison of turbulent inlet boundary conditions between the cases simulated and a case assuming 5\% turbulence intensity.}
     \label{tab:bc}
\end{table}

\begin{figure}[h!]
    \centering
    \includegraphics[width=0.4\textwidth]{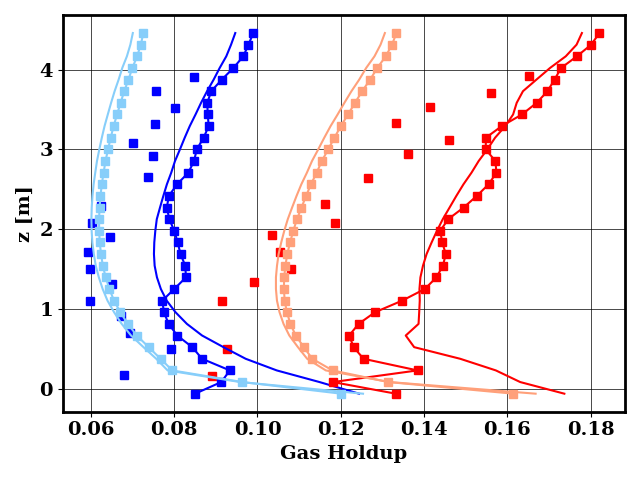}
    \includegraphics[width=0.4\textwidth]{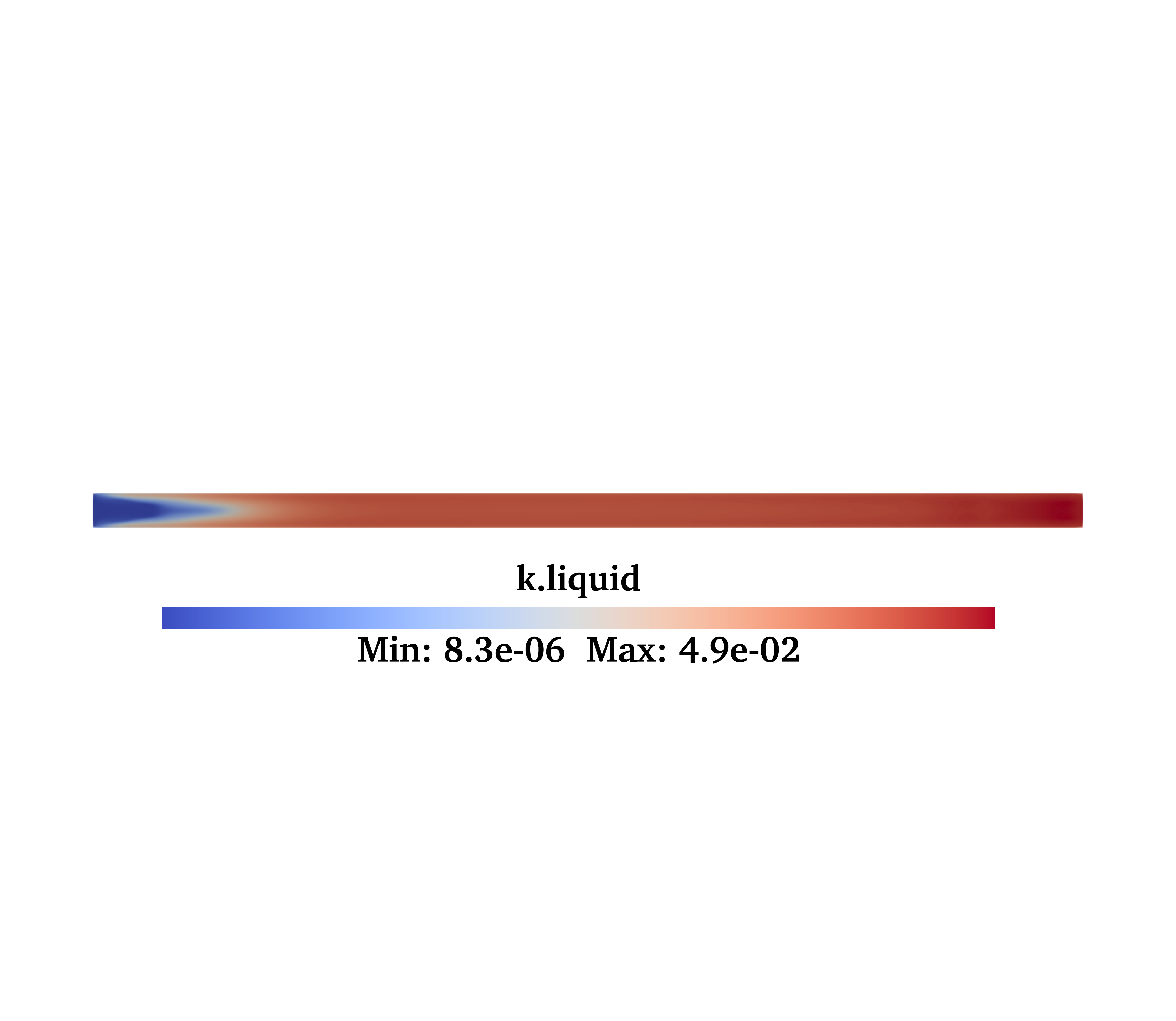}
    \caption{Left: Height conditional average of gas holdup for experiments (\mysquare{black}{black}) \cite{deckwer1978comprehensive}, results using artificially high turbulent inlet boundary conditions (\mythickline{black}), results using freestream turbulence inlet boundary conditions with 5\% turbulence intensity (\mythickbarredsquare{black}{black}) for Exp.~17 (blue) and Exp.~19 (red). Right: contour of the liquid turbulent kinetic energy for Exp.~17 using freestream turbulence inlet boundary conditions with 5\% turbulence intensity and the non-modified binary breakup model \cite{lehr2002bubble}. The flow goes from left to right.}
    \label{fig:correctBC}
\end{figure}

The same numerical simulations as the coarse grid case in Sec.~\ref{sec:disc} were performed with freestream turbulence boundary conditions, assuming 5\% turbulence intensity. The results are shown in Fig.~\ref{fig:correctBC} for gas holdup only for concision. It can be seen that using the non-artificially increased turbulent boundary conditions leads to spurious oscillations for the global breakup model. However, the results are almost unchanged for the binary breakup case. In Fig.~\ref{fig:correctBC} (right) it is clear that the turbulent kinetic energy in the liquid phase rapidly deviates from the freestream boundary conditions because of the walls, and reaches levels higher than the artificially increased turbulent kinetic energy boundary conditions. Therefore, although the inlet boundary conditions chosen are inaccurate, they are unlikely to affect the results reported (aside from providing increased stability). 

\section{\revonerone{Validation of Algo.~\ref{algo:calUnc} and Algo.~\ref{algo:optUnc}}}
\label{app:val}
\revonerone{In this section, we demonstrate a straightforward calibration procedure to validate the ability of Algo.~\ref{algo:calUnc} and Algo.~\ref{algo:optUnc} in accounting for uncertainty due to missing physics.}

\subsection{\revonerone{Simplified bubble size distribution (BSD) dynamics model}}
\revonerone{In this example, we model the dynamics of a bubble size distribution (BSD) within a frozen flow field, i.e., changes in BSD do not influence the fluid flow variables. Each bubble is represented by a particle, which can either merge with another particle (coalescence) or break into multiple particles (breakup). This approach differs from the one used in the main manuscript, where bubble size classes evolve in phase space. The particle-based modeling demonstrated here can be easily extended to a class-based approach by considering a single particle as a group of bubbles. This example is intended solely for illustrative purposes.}

\revonerone{The dynamics are simulated with a timestepping method until a total simulation time is reached. At every timestep, the number of breakup and coalescence events is determined by the breakup and coalescence rates. For each breakup and coalescence event, a random bubble (for breakup) or set of bubbles (for coalescence) is selected to undergo coalescence or breakup. Here, the breakup and coalescence rates are assumed independent of the bubble diameters. While simple, this implementation can result in a breakup or coalescence runaway, especially when breakup and coalescence rates do not balance each other out. In case of breakup runaway,  the number of bubbles grows monotonically, while in coalescence runaway, the number of bubbles decreases monotonically until the simulation ends. The runaway condition is detected as follows: if the number of bubbles falls below 10\% of the initial count (coalescence runaway) or exceeds 10 times the initial count (breakup runaway), runaway is declared, and no BSD is recorded.}
\revonerone{At every breakup or coalescence event, ``$N$-ary" breakup and coalescence occur. If $N=2$, then a binary breakup and binary coalescence are modeled.}

\revonerone{For all the cases, the domain initially contains  $2,000$ bubbles, all with a diameter equal to $1$mm. The timestep is $\Delta t = 0.01$s and the total simulation time is $150$s.}

\revonerone{For each simulation, the BSD is recorded after a statistically stationary solution for the mean bubble diameter is achieved. The statistically stationary state is detected when the window average bubble diameter with a window size of 100 timesteps varies by less than $10$nm. This computational implementation is available in the companion repository.~\footnote{\href{https://github.com/NREL/BioReactorDesign/tree/v0.0.16/tutorial_cases/calibration/bsd/make_dataset}{\revonerone{https://github.com/NREL/BioReactorDesign/tree/v0.0.16/tutorial\_cases/calibration/bsd/make\_dataset}}}}

\subsection{\revonerone{Calibration}}

\revonerone{Two target BSDs are considered: 1) a BSD obtained with a ternary breakup ($N=3$) and coalescence, with rates 0.5 Hz; 2) A BSD obtained with a binary breakup ($N=2$) and coalescence, with a breakup rate $1.6$ Hz and a coalescence rate $2.0$ Hz.}

\revonerone{The numerical model considers binary breakup and coalescence with adjustable breakup and coalescence rates. To avoid breakup or coalescence runaway, the breakup rate is always set close to the coalescence rate. Specifically, the coalescence rate is set in the interval $Cr \in [0.02, 2] Hz$. A breakup rate factor is defined as $Bf \in [0.8, 1.1]$ and the breakup rate is related to the coalescence rate as $Br = Bf Cr$. This ensures that the breakup rates and coalescence rates are close to one another.}

\begin{figure}[h!]
    \centering
    \includegraphics[width=0.5\textwidth]{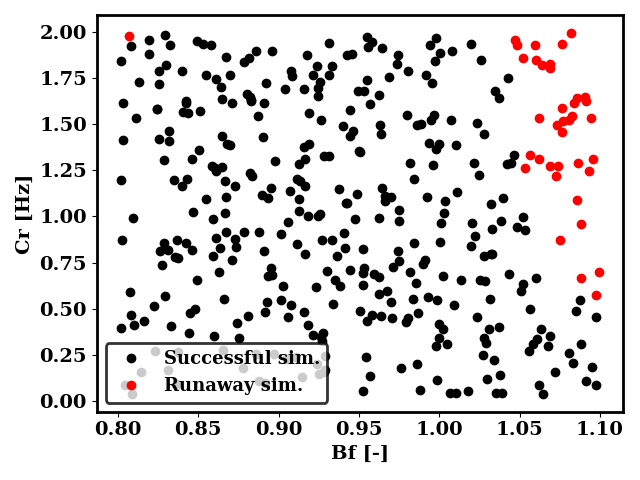}
    \caption{\revonerone{Scatter plot of the simulations run to train a surrogate model of the BSD, including successful (\mythickcircle{black}{black}) and runaway simulations (\mythickcircle{red}{red}).}}
    \label{fig:scatterbsdtoy}
\end{figure}

\revonerone{Figure~\ref{fig:scatterbsdtoy} shows the forward simulations used to construct a dataset for training the surrogate model. The figure also differentiates between stable and runaway simulations. Note that simulations with large or small $Bf$ values lead to runaway events.}

\begin{figure}[h!]
    \centering
    \includegraphics[width=0.45\textwidth]{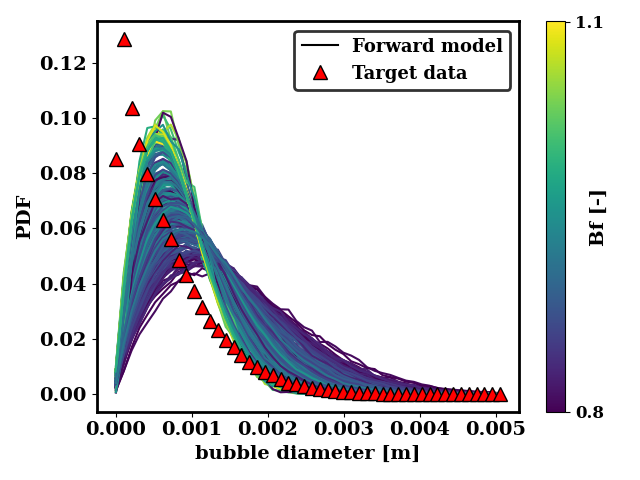}
    \includegraphics[width=0.45\textwidth]{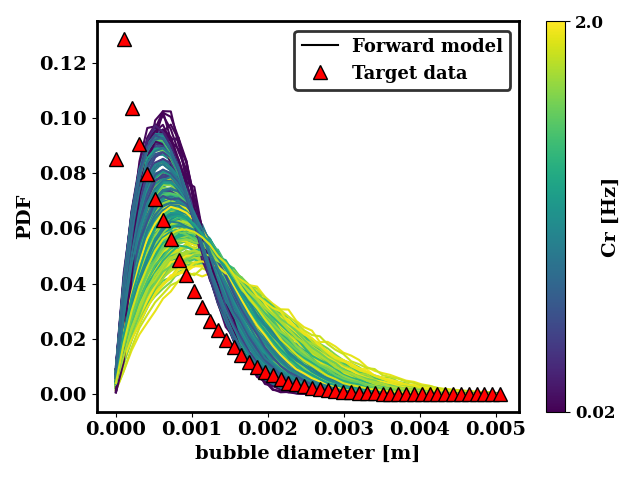}
    \includegraphics[width=0.45\textwidth]{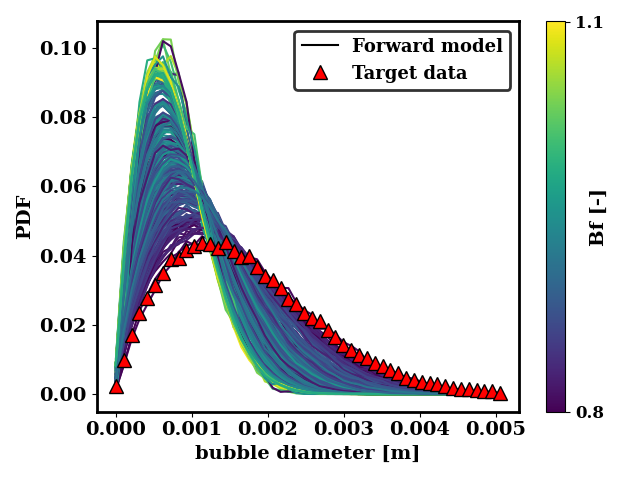}
    \includegraphics[width=0.45\textwidth]{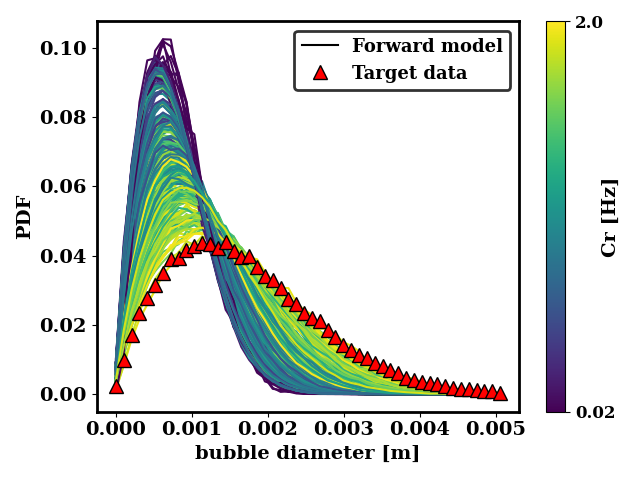}
    \caption{\revonerone{Forward model prediction of BSD colored by $Bf$ (left) and $Cr$ (right) for the ternary breakup and coalescence target (top) and the binary breakup and coalescence target (bottom).}}
    \label{fig:target-sim}
\end{figure}

\revonerone{Figure~\ref{fig:target-sim} compares the target data with the simulated data. It is clear that for ternary breakup, the training data generated missed a significant amount of physics, while for binary breakup, the training data aligns more closely to the target. This example illustrates a scenario where a numerical model's predictions are inconsistent with observations, possibly due to the simulation's inability to model a specific physical phenomenon. The rest of this section, verifies that  Algo.~\ref{algo:calUnc} and Algo.~\ref{algo:optUnc} predict that ternary breakup and coalescence models requires a larger missing physics uncertainty compared to the binary breakup and coalescence models.}

\revonerone{Using 400 BSD numerically generated (shown in Fig.~\ref{fig:scatterbsdtoy}), a neural network surrogate model is constructed similar to Fig.~\ref{fig:nnstruct} with 3 inputs ($Bf$, $Cr$, and the bubble diameter), with 4 hidden layers of size $\{20,20,20,10\}$, a batch size of $128$, $6000$ epochs and a learning rate of $2\times 10^{-4}$. The surrogate model is subsequently used to calibrate the $Bf$ and $Cr$.} 

\begin{figure}[h!]
    \centering
    \includegraphics[width=0.33\textwidth]{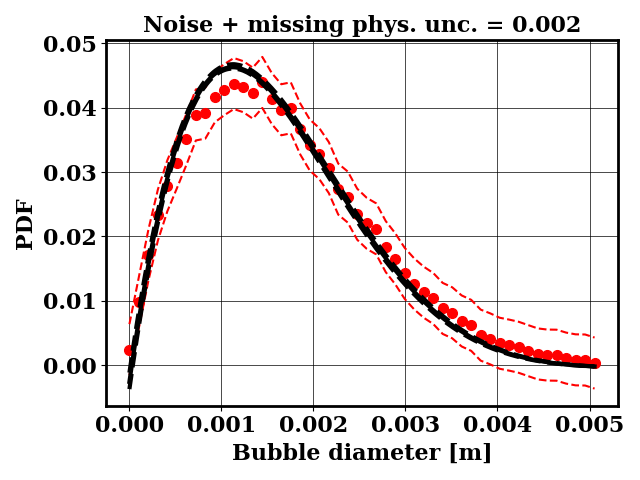}
    \includegraphics[width=0.33\textwidth]{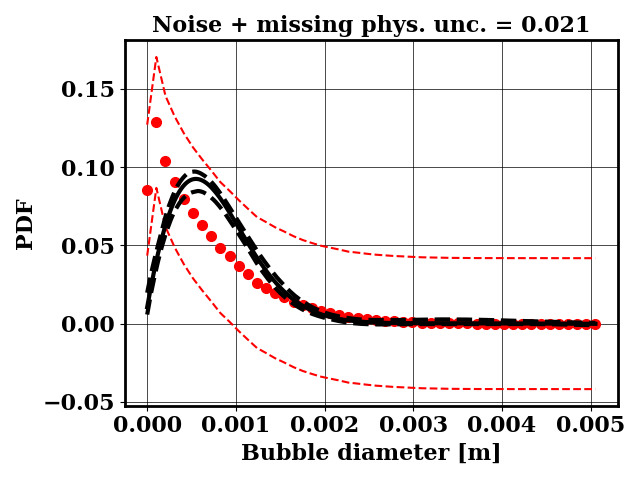}
    \includegraphics[width=0.33\textwidth]{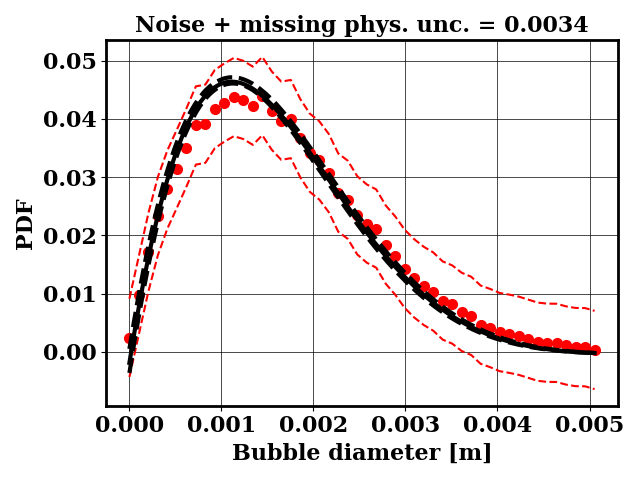}
    \includegraphics[width=0.33\textwidth]{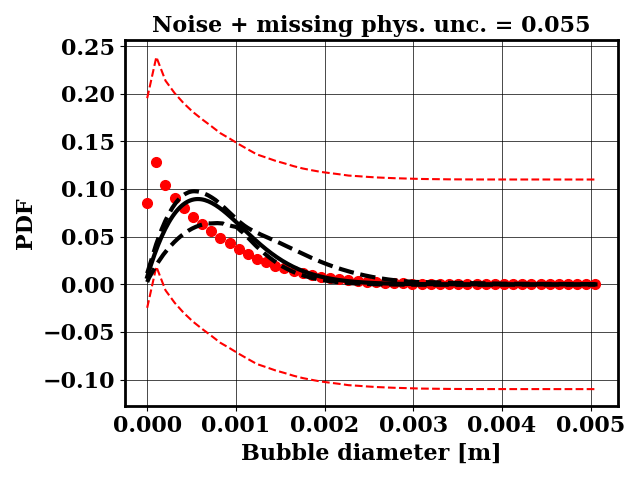}
    \caption{\revonerone{Mean (\mythickline{black}) and 95\% confidence interval (\mythickdashedline{black}) of the normal distribution model. Predictions are overlayed with target observations (\mythickcircle{red}{red}) and the corresponding missing physics and noise uncertainty (\mythickdashedline{red}). Top: results of Algo.~\ref{algo:calUnc}. Bottom: results of Algo.~\ref{algo:optUnc}. Left: binary breakup and coalescence target. Right: ternary breakup and coalescence target.}}
    \label{fig:simplecase}
\end{figure}

\revonerone{The prior PDFs of $Bf$ and $Cr$ are uniformly distributed as $\mathcal{U}(0.8, 1.1)$ and $\mathcal{U}(0.02, 2)$ respectively. For Algo.~\ref{algo:calUnc}, the prior distribution of the missing physics uncertainty is uniformly distributed as $\mathcal{U}(0.001, 0.1)$. Similar to Sec.~\ref{sec:postpar}, the calibration is made using $10,000$ warmup samples and $2,000$ posterior samples using Hamiltonian Monte-Carlo (HMC) \cite{hoffman2014no, numpyro}. The propagated uncertainty of the calibrated parameters (akin to Fig.~\ref{fig:missPhy}) is shown in Fig.~\ref{fig:simplecase}.}

\begin{figure}[h!]
    \centering
    \includegraphics[width=0.33\textwidth]{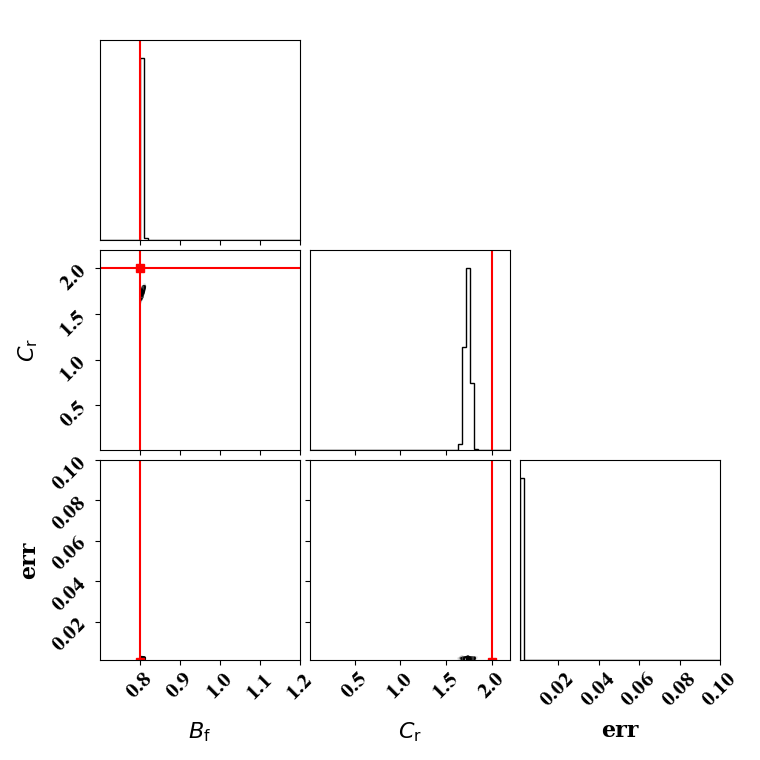}
    \includegraphics[width=0.33\textwidth]{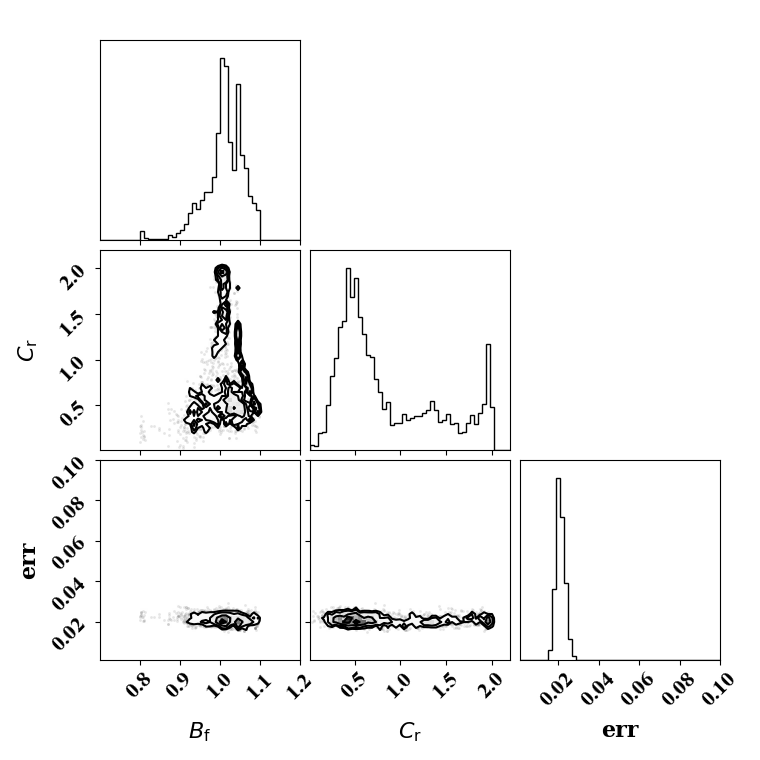}
    \includegraphics[width=0.33\textwidth]{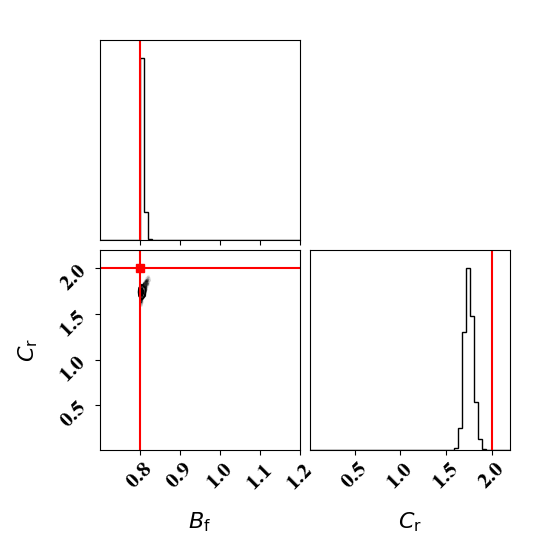}
    \includegraphics[width=0.33\textwidth]{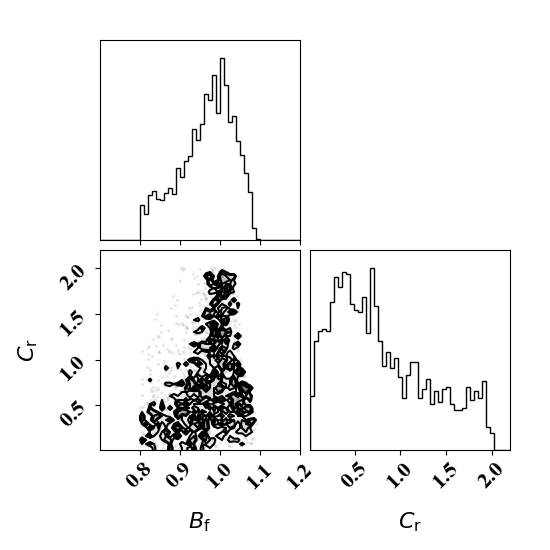}
    \caption{\revonerone{PDF of the calibrated $Bf$ and $Cr$ parameters. Predictions are overlayed with true values for the binary breakup and coalescence target (\mythickline{red}). Top: results of Algo.~\ref{algo:calUnc}. Bottom: results of Algo.~\ref{algo:optUnc}. Left: binary breakup and coalescence target. Right: ternary breakup and coalescence target.}}
    \label{fig:cal_simplecase}
\end{figure}

\revonerone{Note that the noise in the target data is similar between both targets since it is only due to the statistical uncertainty. Figure~\ref{fig:simplecase} shows that Algo.~\ref{algo:calUnc} and Algo.~\ref{algo:optUnc} predict a significantly lower missing physics uncertainty for the second target (binary breakup and coalescence) than for the first target (ternary breakup and coalescence). Similar to Tab.~\ref{tab:likePhy}, Algo.~\ref{algo:optUnc} tends to lead to higher likelihood uncertainties. The calibrated parameters are shown in Fig.~\ref{fig:cal_simplecase}, showing that calibration results are consistent between Algo.~\ref{algo:calUnc} and Algo.~\ref{algo:optUnc}.}

\section{\revone{Effect of calibration on the BSD}}
\label{app:effectBSD}
\begin{figure}[h!]
    \centering
    \includegraphics[width=0.3\textwidth]{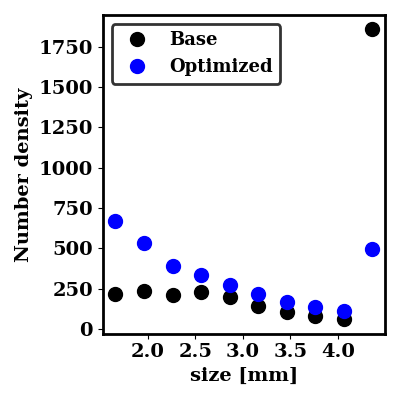}
    \caption{\revone{Bubble size distribution at $z=1$m for experiment 17 and the global breakup model \cite{laakkonen2007modelling} of the base model (\mycircle{black}{black}) and the optimized model (\mycircle{blue}{blue}).}}
    \label{fig:bsdopt}
\end{figure}

\revone{Although the bubble size itself is not a variable that is used as a calibration target, the effect of the calibration on the BSD can be evaluated. In Fig.~\ref{fig:bsdopt}, the BSD is plotted at $z=1$m for experiment 17 simulated with the global breakup model. It is clear that in the base model, bubbles rapidly coalesce and saturate the highest bin size modeled for the BSD, while the issue is less potent in the calibrated model. This is a direct effect of the larger breakup rate in the calibrated case.}

\revone{Even in the calibrated cases, the definition of the bounds of the bubble size affected the BSD. Given that the calibration results already suggest a larger breakup rate, removing this effect might even amplify the magnitude of the efficiency factor correction, and might help bridge the gap to the stirred-tank reactor BSD measurements (Fig.~\ref{fig:bsd}). This investigation is left for future work.}

\section{\revtwo{Calibrated parameter sensitivity}}
\label{app:sens}
\revtwo{The posterior distributions of calibrated parameters (Sec.~\ref{sec:postpar}) suggest that it is not possible to conclude about the value of surface tension based on the experimental dataset. On the one hand, if surface tension is inaccurate, the calibration procedure cannot identify it. On the other hand, if the objective is to capture gas hold up and \ce{CO2} mole fraction, any surface tension guess is reasonable (within the support of the prior distribution).}

\begin{figure}[h!]
    \centering
    \includegraphics[width=0.43\textwidth]{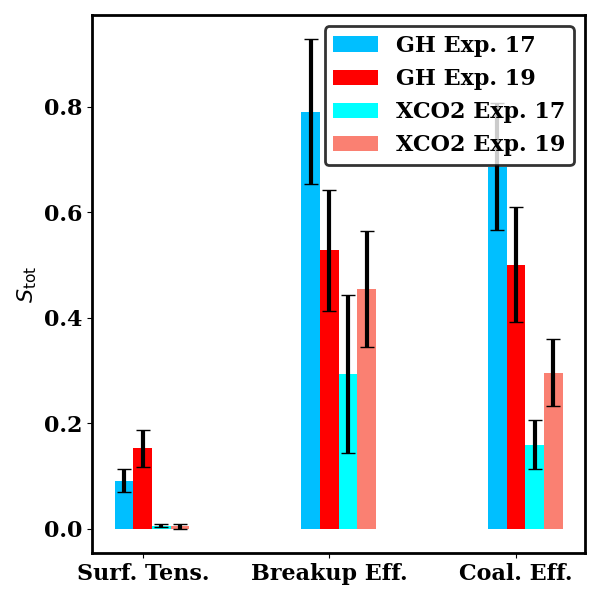}
    \includegraphics[width=0.43\textwidth]{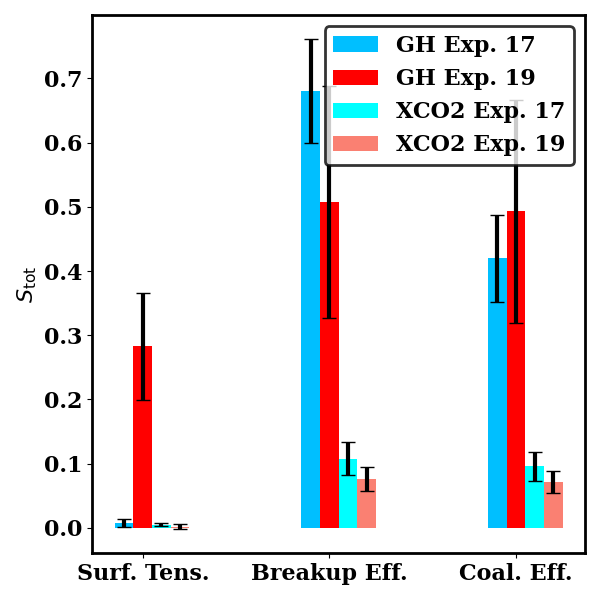}
    \caption{\revtwo{Total Sobol indices averaged over 100 height values for all four experimental datasets. Error bar denote standard deviation of total Sobol indices across different heights in the reactor. Left: global breakup model \cite{laakkonen2007modelling}. Right: binary breakup model \cite{lehr2002bubble}.}}
    \label{fig:sensitivity}
\end{figure}

\revtwo{To confirm that surface tension is less impactful than the other calibrated parameters, a sensitivity analysis was conducted by computing the total Sobol indices for gas holdup and \ce{CO2} mole fraction. The Sobol indices are computed using $2^{16}$ evaluations of the trained surrogate model for 100 different heights ($z$) sampled uniformly from the lowest to the highest height available in the dataset \cite{saltelli2002making}. Figure~\ref{fig:sensitivity} shows the average total Sobol indices across all heights, for all three calibrated parameters, and all four experimental datasets. It is clear that in all cases, gas holdup and \ce{CO2} mole fraction are less sensitive to surface tension than breakup and coalescence efficiency factors.}

\begin{thebibliography}{86}
\expandafter\ifx\csname natexlab\endcsname\relax\def\natexlab#1{#1}\fi
\providecommand{\url}[1]{\texttt{#1}}
\providecommand{\href}[2]{#2}
\providecommand{\path}[1]{#1}
\providecommand{\DOIprefix}{doi:}
\providecommand{\ArXivprefix}{arXiv:}
\providecommand{\URLprefix}{URL: }
\providecommand{\Pubmedprefix}{pmid:}
\providecommand{\doi}[1]{\href{http://dx.doi.org/#1}{\path{#1}}}
\providecommand{\Pubmed}[1]{\href{pmid:#1}{\path{#1}}}
\providecommand{\bibinfo}[2]{#2}
\ifx\xfnm\relax \def\xfnm[#1]{\unskip,\space#1}\fi
\bibitem[{Lee et~al.(2023)Lee, Calvin, Dasgupta, Krinner, Mukherji, Thorne, Trisos, Romero, Aldunce, Barret et~al.}]{lee2023ipcc}
\bibinfo{author}{H.~Lee}, \bibinfo{author}{K.~Calvin}, \bibinfo{author}{D.~Dasgupta}, \bibinfo{author}{G.~Krinner}, \bibinfo{author}{A.~Mukherji}, \bibinfo{author}{P.~Thorne}, \bibinfo{author}{C.~Trisos}, \bibinfo{author}{J.~Romero}, \bibinfo{author}{P.~Aldunce}, \bibinfo{author}{K.~Barret}, et~al.,
\newblock \bibinfo{title}{{IPCC, 2023: Climate Change 2023: Synthesis Report, Summary for Policymakers. Contribution of Working Groups I, II and III to the Sixth Assessment Report of the Intergovernmental Panel on Climate Change}}  (\bibinfo{year}{2023}).
\bibitem[{Bergero et~al.(2023)Bergero, Gosnell, Gielen, Kang, Bazilian, and Davis}]{bergero2023pathways}
\bibinfo{author}{C.~Bergero}, \bibinfo{author}{G.~Gosnell}, \bibinfo{author}{D.~Gielen}, \bibinfo{author}{S.~Kang}, \bibinfo{author}{M.~Bazilian}, \bibinfo{author}{S.~J. Davis},
\newblock \bibinfo{title}{Pathways to net-zero emissions from aviation},
\newblock \bibinfo{journal}{Nature Sustainability} \bibinfo{volume}{6} (\bibinfo{year}{2023}) \bibinfo{pages}{404--414}.
\bibitem[{Schwab et~al.(2021)Schwab, Thomas, Bennett, Robertson, and Cary}]{schwab2021electrification}
\bibinfo{author}{A.~Schwab}, \bibinfo{author}{A.~Thomas}, \bibinfo{author}{J.~Bennett}, \bibinfo{author}{E.~Robertson}, \bibinfo{author}{S.~Cary}, \bibinfo{title}{{Electrification of aircraft: Challenges, barriers, and potential impacts}}, \bibinfo{type}{Technical Report}, National Renewable Energy Lab.(NREL), Golden, CO (United States), \bibinfo{year}{2021}.
\bibitem[{Shahriar and Khanal(2022)}]{shahriar2022current}
\bibinfo{author}{M.~F. Shahriar}, \bibinfo{author}{A.~Khanal},
\newblock \bibinfo{title}{{The current techno-economic, environmental, policy status and perspectives of sustainable aviation fuel (SAF)}},
\newblock \bibinfo{journal}{Fuel} \bibinfo{volume}{325} (\bibinfo{year}{2022}) \bibinfo{pages}{124905}.
\bibitem[{Grim et~al.(2023)Grim, Ferrell~III, Huang, Tao, and Resch}]{grim2023feasibility}
\bibinfo{author}{R.~G. Grim}, \bibinfo{author}{J.~R. Ferrell~III}, \bibinfo{author}{Z.~Huang}, \bibinfo{author}{L.~Tao}, \bibinfo{author}{M.~G. Resch},
\newblock \bibinfo{title}{{The feasibility of direct CO2 conversion technologies on impacting mid-century climate goals}},
\newblock \bibinfo{journal}{Joule} \bibinfo{volume}{7} (\bibinfo{year}{2023}) \bibinfo{pages}{1684--1699}.
\bibitem[{Tarraran et~al.(2023)Tarraran, Agostino, Vasile, Azim, Antonicelli, Baker, Millard, Re, Menin, Tommasi et~al.}]{tarraran2023high}
\bibinfo{author}{L.~Tarraran}, \bibinfo{author}{V.~Agostino}, \bibinfo{author}{N.~S. Vasile}, \bibinfo{author}{A.~A. Azim}, \bibinfo{author}{G.~Antonicelli}, \bibinfo{author}{J.~Baker}, \bibinfo{author}{J.~Millard}, \bibinfo{author}{A.~Re}, \bibinfo{author}{B.~Menin}, \bibinfo{author}{T.~Tommasi}, et~al.,
\newblock \bibinfo{title}{{High-pressure fermentation of CO2 and H2 by a modified Acetobacterium woodii}},
\newblock \bibinfo{journal}{Journal of CO2 Utilization} \bibinfo{volume}{76} (\bibinfo{year}{2023}) \bibinfo{pages}{102583}.
\bibitem[{Straub et~al.(2014)Straub, Demler, Weuster-Botz, and D{\"u}rre}]{straub2014selective}
\bibinfo{author}{M.~Straub}, \bibinfo{author}{M.~Demler}, \bibinfo{author}{D.~Weuster-Botz}, \bibinfo{author}{P.~D{\"u}rre},
\newblock \bibinfo{title}{{Selective enhancement of autotrophic acetate production with genetically modified Acetobacterium woodii}},
\newblock \bibinfo{journal}{Journal of Biotechnology} \bibinfo{volume}{178} (\bibinfo{year}{2014}) \bibinfo{pages}{67--72}.
\bibitem[{Pfeifer et~al.(2021)Pfeifer, Ergal, Koller, Basen, Schuster, and Simon}]{pfeifer2021archaea}
\bibinfo{author}{K.~Pfeifer}, \bibinfo{author}{{\.I}.~Ergal}, \bibinfo{author}{M.~Koller}, \bibinfo{author}{M.~Basen}, \bibinfo{author}{B.~Schuster}, \bibinfo{author}{K.-M.~R. Simon},
\newblock \bibinfo{title}{Archaea biotechnology},
\newblock \bibinfo{journal}{Biotechnology Advances} \bibinfo{volume}{47} (\bibinfo{year}{2021}) \bibinfo{pages}{107668}.
\bibitem[{Yan et~al.(2023)Yan, Wang, Zhu, Zhang, and Luo}]{yan2023mechanisms}
\bibinfo{author}{S.-l. Yan}, \bibinfo{author}{X.-q. Wang}, \bibinfo{author}{L.-t. Zhu}, \bibinfo{author}{X.-b. Zhang}, \bibinfo{author}{Z.-h. Luo},
\newblock \bibinfo{title}{Mechanisms and modeling of bubble dynamic behaviors and mass transfer under gravity: a review},
\newblock \bibinfo{journal}{Chemical Engineering Science}  (\bibinfo{year}{2023}) \bibinfo{pages}{118854}.
\bibitem[{McGraw(1997)}]{mcgraw1997description}
\bibinfo{author}{R.~McGraw},
\newblock \bibinfo{title}{Description of aerosol dynamics by the quadrature method of moments},
\newblock \bibinfo{journal}{Aerosol Science and Technology} \bibinfo{volume}{27} (\bibinfo{year}{1997}) \bibinfo{pages}{255--265}.
\bibitem[{Krepper et~al.(2007)Krepper, Frank, Lucas, Prasser, and Zwart}]{krepper2007inhomogeneous}
\bibinfo{author}{E.~Krepper}, \bibinfo{author}{T.~Frank}, \bibinfo{author}{D.~Lucas}, \bibinfo{author}{H.-M. Prasser}, \bibinfo{author}{P.~J. Zwart},
\newblock \bibinfo{title}{{Inhomogeneous MUSIG model - a population balance approach for polydispersed bubbly flows}},
\newblock in: \bibinfo{booktitle}{The 12th International Topical Meeting on Nuclear Reactor Thermal Hydraulics (NURETH-12)}, volume~\bibinfo{volume}{30}, \bibinfo{year}{2007}, p. \bibinfo{pages}{2007}.
\bibitem[{Lo(1996)}]{lo1996application}
\bibinfo{author}{S.~Lo}, \bibinfo{title}{{Application of population balance to CFD modeling of bubbly flow via the MUSIG model}}, \bibinfo{year}{1996}.
\bibitem[{Marchisio et~al.(2003)Marchisio, Pikturna, Fox, Vigil, and Barresi}]{marchisio2003quadrature}
\bibinfo{author}{D.~L. Marchisio}, \bibinfo{author}{J.~T. Pikturna}, \bibinfo{author}{R.~O. Fox}, \bibinfo{author}{R.~D. Vigil}, \bibinfo{author}{A.~A. Barresi},
\newblock \bibinfo{title}{Quadrature method of moments for population-balance equations},
\newblock \bibinfo{journal}{AIChE Journal} \bibinfo{volume}{49} (\bibinfo{year}{2003}) \bibinfo{pages}{1266--1276}.
\bibitem[{Gao et~al.(2016)Gao, Li, Buffo, Podg{\'o}rska, and Marchisio}]{gao2016simulation}
\bibinfo{author}{Z.~Gao}, \bibinfo{author}{D.~Li}, \bibinfo{author}{A.~Buffo}, \bibinfo{author}{W.~Podg{\'o}rska}, \bibinfo{author}{D.~L. Marchisio},
\newblock \bibinfo{title}{{Simulation of droplet breakage in turbulent liquid--liquid dispersions with CFD-PBM: Comparison of breakage kernels}},
\newblock \bibinfo{journal}{Chemical Engineering Science} \bibinfo{volume}{142} (\bibinfo{year}{2016}) \bibinfo{pages}{277--288}.
\bibitem[{K{\'a}lal et~al.(2014)K{\'a}lal, Jahoda, and Fo{\v{r}}t}]{kalal2014modelling}
\bibinfo{author}{Z.~K{\'a}lal}, \bibinfo{author}{M.~Jahoda}, \bibinfo{author}{I.~Fo{\v{r}}t},
\newblock \bibinfo{title}{{Modelling of the bubble size distribution in an aerated stirred tank: Theoretical and numerical comparison of different breakup models}},
\newblock \bibinfo{journal}{Chemical and Process Engineering}  (\bibinfo{year}{2014}) \bibinfo{pages}{331--348}.
\bibitem[{Wang et~al.(2005)Wang, Wang, and Jin}]{wang2005population}
\bibinfo{author}{T.~Wang}, \bibinfo{author}{J.~Wang}, \bibinfo{author}{Y.~Jin},
\newblock \bibinfo{title}{{Population balance model for gas- liquid flows: Influence of bubble coalescence and breakup models}},
\newblock \bibinfo{journal}{Industrial \& Engineering Chemistry Research} \bibinfo{volume}{44} (\bibinfo{year}{2005}) \bibinfo{pages}{7540--7549}.
\bibitem[{Li and Liao(2024)}]{li2024cfd}
\bibinfo{author}{S.~Li}, \bibinfo{author}{Y.~Liao},
\newblock \bibinfo{title}{{CFD investigation of bubble breakup and coalescence in a rectangular pool-scrubbing column}},
\newblock \bibinfo{journal}{Nuclear Engineering and Design} \bibinfo{volume}{425} (\bibinfo{year}{2024}) \bibinfo{pages}{113342}.
\bibitem[{Laakkonen et~al.(2006)Laakkonen, Alopaeus, and Aittamaa}]{laakkonen2006validation}
\bibinfo{author}{M.~Laakkonen}, \bibinfo{author}{V.~Alopaeus}, \bibinfo{author}{J.~Aittamaa},
\newblock \bibinfo{title}{Validation of bubble breakage, coalescence and mass transfer models for gas--liquid dispersion in agitated vessel},
\newblock \bibinfo{journal}{Chemical Engineering Science} \bibinfo{volume}{61} (\bibinfo{year}{2006}) \bibinfo{pages}{218--228}.
\bibitem[{Laakkonen et~al.(2007)Laakkonen, Moilanen, Alopaeus, and Aittamaa}]{laakkonen2007modelling}
\bibinfo{author}{M.~Laakkonen}, \bibinfo{author}{P.~Moilanen}, \bibinfo{author}{V.~Alopaeus}, \bibinfo{author}{J.~Aittamaa},
\newblock \bibinfo{title}{Modelling local bubble size distributions in agitated vessels},
\newblock \bibinfo{journal}{Chemical Engineering Science} \bibinfo{volume}{62} (\bibinfo{year}{2007}) \bibinfo{pages}{721--740}.
\bibitem[{Singh et~al.(2009)Singh, Mahajani, Shenoy, and Ghosh}]{singh2009population}
\bibinfo{author}{K.~Singh}, \bibinfo{author}{S.~Mahajani}, \bibinfo{author}{K.~Shenoy}, \bibinfo{author}{S.~Ghosh},
\newblock \bibinfo{title}{Population balance modeling of liquid- liquid dispersions in homogeneous continuous-flow stirred tank},
\newblock \bibinfo{journal}{Industrial \& Engineering Chemistry research} \bibinfo{volume}{48} (\bibinfo{year}{2009}) \bibinfo{pages}{8121--8133}.
\bibitem[{Gel et~al.(2023)Gel, Vaidheeswaran, and Clarke}]{gel2023comparison}
\bibinfo{author}{A.~Gel}, \bibinfo{author}{A.~Vaidheeswaran}, \bibinfo{author}{M.~A. Clarke},
\newblock \bibinfo{title}{{Comparison of Deterministic and Bayesian Calibration of MFiX-PIC, Part 1: Settling Bed}},
\newblock \bibinfo{journal}{arXiv preprint arXiv:2305.01132}  (\bibinfo{year}{2023}).
\bibitem[{Deckwer et~al.(1978)Deckwer, Adler, and Zaidi}]{deckwer1978comprehensive}
\bibinfo{author}{W.-D. Deckwer}, \bibinfo{author}{I.~Adler}, \bibinfo{author}{A.~Zaidi},
\newblock \bibinfo{title}{{A comprehensive study on CO2-interphase mass transfer in vertical cocurrent and countercurrent gas-liquid flow}},
\newblock \bibinfo{journal}{The Canadian Journal of Chemical Engineering} \bibinfo{volume}{56} (\bibinfo{year}{1978}) \bibinfo{pages}{43--55}.
\bibitem[{Alopaeus et~al.(2002)Alopaeus, Koskinen, Keskinen, and Majander}]{alopaeus2002simulation}
\bibinfo{author}{V.~Alopaeus}, \bibinfo{author}{J.~Koskinen}, \bibinfo{author}{K.~I. Keskinen}, \bibinfo{author}{J.~Majander},
\newblock \bibinfo{title}{{Simulation of the population balances for liquid--liquid systems in a nonideal stirred tank. Part 2—parameter fitting and the use of the multiblock model for dense dispersions}},
\newblock \bibinfo{journal}{Chemical Engineering Science} \bibinfo{volume}{57} (\bibinfo{year}{2002}) \bibinfo{pages}{1815--1825}.
\bibitem[{Ruiz and Padilla(2005)}]{ruiz2005determination}
\bibinfo{author}{M.~Ruiz}, \bibinfo{author}{R.~Padilla},
\newblock \bibinfo{title}{Determination of coalescence functions in liquid--liquid dispersions},
\newblock \bibinfo{journal}{Hydrometallurgy} \bibinfo{volume}{80} (\bibinfo{year}{2005}) \bibinfo{pages}{32--42}.
\bibitem[{Azizi and Al~Taweel(2011)}]{azizi2011turbulently}
\bibinfo{author}{F.~Azizi}, \bibinfo{author}{A.~Al~Taweel},
\newblock \bibinfo{title}{{Turbulently flowing liquid--liquid dispersions. Part I: drop breakage and coalescence}},
\newblock \bibinfo{journal}{Chemical Engineering Journal} \bibinfo{volume}{166} (\bibinfo{year}{2011}) \bibinfo{pages}{715--725}.
\bibitem[{Castellano et~al.(2019)Castellano, Carrillo, Sheibat-Othman, Marchisio, Buffo, and Charton}]{castellano2019using}
\bibinfo{author}{S.~Castellano}, \bibinfo{author}{L.~Carrillo}, \bibinfo{author}{N.~Sheibat-Othman}, \bibinfo{author}{D.~Marchisio}, \bibinfo{author}{A.~Buffo}, \bibinfo{author}{S.~Charton},
\newblock \bibinfo{title}{{Using the full turbulence spectrum for describing droplet coalescence and breakage in industrial liquid-liquid systems: Experiments and modeling}},
\newblock \bibinfo{journal}{Chemical Engineering Journal} \bibinfo{volume}{374} (\bibinfo{year}{2019}) \bibinfo{pages}{1420--1432}.
\bibitem[{Sathyagal et~al.(1995)Sathyagal, Ramkrishna, and Narsimhan}]{sathyagal1995solution}
\bibinfo{author}{A.~Sathyagal}, \bibinfo{author}{D.~Ramkrishna}, \bibinfo{author}{G.~Narsimhan},
\newblock \bibinfo{title}{{Solution of inverse problems in population balances-II. Particle break-up}},
\newblock \bibinfo{journal}{Computers \& Chemical Engineering} \bibinfo{volume}{19} (\bibinfo{year}{1995}) \bibinfo{pages}{437--451}.
\bibitem[{Mignard et~al.(2006)Mignard, Amin, and Ni}]{mignard2006determination}
\bibinfo{author}{D.~Mignard}, \bibinfo{author}{L.~P. Amin}, \bibinfo{author}{X.~Ni},
\newblock \bibinfo{title}{Determination of breakage rates of oil droplets in a continuous oscillatory baffled tube},
\newblock \bibinfo{journal}{Chemical Engineering Science} \bibinfo{volume}{61} (\bibinfo{year}{2006}) \bibinfo{pages}{6902--6917}.
\bibitem[{Maluta et~al.(2021)Maluta, Buffo, Marchisio, Montante, Paglianti, and Vanni}]{maluta2021effect}
\bibinfo{author}{F.~Maluta}, \bibinfo{author}{A.~Buffo}, \bibinfo{author}{D.~Marchisio}, \bibinfo{author}{G.~Montante}, \bibinfo{author}{A.~Paglianti}, \bibinfo{author}{M.~Vanni},
\newblock \bibinfo{title}{Effect of turbulent kinetic energy dissipation rate on the prediction of droplet size distribution in stirred tanks},
\newblock \bibinfo{journal}{International Journal of Multiphase Flow} \bibinfo{volume}{136} (\bibinfo{year}{2021}) \bibinfo{pages}{103547}.
\bibitem[{Maa{\ss} and Kraume(2012)}]{maass2012determination}
\bibinfo{author}{S.~Maa{\ss}}, \bibinfo{author}{M.~Kraume},
\newblock \bibinfo{title}{Determination of breakage rates using single drop experiments},
\newblock \bibinfo{journal}{Chemical Engineering Science} \bibinfo{volume}{70} (\bibinfo{year}{2012}) \bibinfo{pages}{146--164}.
\bibitem[{Solsvik et~al.(2014)Solsvik, Becker, Sheibat-Othman, and Jakobsen}]{solsvik2014population}
\bibinfo{author}{J.~Solsvik}, \bibinfo{author}{P.~J. Becker}, \bibinfo{author}{N.~Sheibat-Othman}, \bibinfo{author}{H.~A. Jakobsen},
\newblock \bibinfo{title}{{Population balance model: Breakage kernel parameter estimation to emulsification data}},
\newblock \bibinfo{journal}{The Canadian Journal of Chemical Engineering} \bibinfo{volume}{92} (\bibinfo{year}{2014}) \bibinfo{pages}{1082--1099}.
\bibitem[{Becker et~al.(2014)Becker, Puel, Jakobsen, and Sheibat-Othman}]{becker2014development}
\bibinfo{author}{P.~J. Becker}, \bibinfo{author}{F.~Puel}, \bibinfo{author}{H.~A. Jakobsen}, \bibinfo{author}{N.~Sheibat-Othman},
\newblock \bibinfo{title}{Development of an improved breakage kernel for high dispersed viscosity phase emulsification},
\newblock \bibinfo{journal}{Chemical Engineering Science} \bibinfo{volume}{109} (\bibinfo{year}{2014}) \bibinfo{pages}{326--338}.
\bibitem[{Rzehak et~al.(2017)Rzehak, Ziegenhein, Kriebitzsch, Krepper, and Lucas}]{rzehak2017unified}
\bibinfo{author}{R.~Rzehak}, \bibinfo{author}{T.~Ziegenhein}, \bibinfo{author}{S.~Kriebitzsch}, \bibinfo{author}{E.~Krepper}, \bibinfo{author}{D.~Lucas},
\newblock \bibinfo{title}{Unified modeling of bubbly flows in pipes, bubble columns, and airlift columns},
\newblock \bibinfo{journal}{Chemical Engineering Science} \bibinfo{volume}{157} (\bibinfo{year}{2017}) \bibinfo{pages}{147--158}.
\bibitem[{Besagni et~al.(2017)Besagni, Inzoli, Ziegenhein, and Lucas}]{besagni2017computational}
\bibinfo{author}{G.~Besagni}, \bibinfo{author}{F.~Inzoli}, \bibinfo{author}{T.~Ziegenhein}, \bibinfo{author}{D.~Lucas},
\newblock \bibinfo{title}{Computational fluid-dynamic modeling of the pseudo-homogeneous flow regime in large-scale bubble columns},
\newblock \bibinfo{journal}{Chemical Engineering Science} \bibinfo{volume}{160} (\bibinfo{year}{2017}) \bibinfo{pages}{144--160}.
\bibitem[{Tierney(1994)}]{tierney1994markov}
\bibinfo{author}{L.~Tierney},
\newblock \bibinfo{title}{Markov chains for exploring posterior distributions},
\newblock \bibinfo{journal}{Ann. Stat.}  (\bibinfo{year}{1994}) \bibinfo{pages}{1701--1728}.
\bibitem[{Roberts and Smith(1994)}]{roberts1994simple}
\bibinfo{author}{G.~Roberts}, \bibinfo{author}{A.~Smith},
\newblock \bibinfo{title}{{Simple conditions for the convergence of the Gibbs sampler and Metropolis-Hastings algorithms}},
\newblock \bibinfo{journal}{Stochastic Process. Appl.} \bibinfo{volume}{49} (\bibinfo{year}{1994}) \bibinfo{pages}{207--216}.
\bibitem[{Gelman et~al.(1997)Gelman, Gilks, and Roberts}]{gelman1997weak}
\bibinfo{author}{A.~Gelman}, \bibinfo{author}{W.~Gilks}, \bibinfo{author}{G.~Roberts},
\newblock \bibinfo{title}{{Weak convergence and optimal scaling of random walk Metropolis algorithms}},
\newblock \bibinfo{journal}{Ann. Appl. Probab.} \bibinfo{volume}{7} (\bibinfo{year}{1997}) \bibinfo{pages}{110--120}.
\bibitem[{Braman et~al.(2013)Braman, Oliver, and Raman}]{braman2013bayesian}
\bibinfo{author}{K.~Braman}, \bibinfo{author}{T.~A. Oliver}, \bibinfo{author}{V.~Raman},
\newblock \bibinfo{title}{Bayesian analysis of syngas chemistry models},
\newblock \bibinfo{journal}{Combustion Theory and Modelling} \bibinfo{volume}{17} (\bibinfo{year}{2013}) \bibinfo{pages}{858--887}.
\bibitem[{Hassanaly et~al.(2021)Hassanaly, Sitaraman, Schulte, Ptak, Simon, Udwary, Leach, and Splawn}]{hassanaly2021surface}
\bibinfo{author}{M.~Hassanaly}, \bibinfo{author}{H.~Sitaraman}, \bibinfo{author}{K.~L. Schulte}, \bibinfo{author}{A.~J. Ptak}, \bibinfo{author}{J.~Simon}, \bibinfo{author}{K.~Udwary}, \bibinfo{author}{J.~H. Leach}, \bibinfo{author}{H.~Splawn},
\newblock \bibinfo{title}{Surface chemistry models for gaas epitaxial growth and hydride cracking using reacting flow simulations},
\newblock \bibinfo{journal}{Journal of Applied Physics} \bibinfo{volume}{130} (\bibinfo{year}{2021}).
\bibitem[{Bell et~al.(2019)Bell, Day, Goodman, Grout, and Morzfeld}]{bell2019bayesian}
\bibinfo{author}{J.~Bell}, \bibinfo{author}{M.~Day}, \bibinfo{author}{J.~Goodman}, \bibinfo{author}{R.~Grout}, \bibinfo{author}{M.~Morzfeld},
\newblock \bibinfo{title}{{A Bayesian approach to calibrating hydrogen flame kinetics using many experiments and parameters}},
\newblock \bibinfo{journal}{Combustion and Flame} \bibinfo{volume}{205} (\bibinfo{year}{2019}) \bibinfo{pages}{305--315}.
\bibitem[{Hassanaly et~al.(2024)Hassanaly, Weddle, King, De, Doostan, Randall, Dufek, Colclasure, and Smith}]{hassanaly2023pinn2}
\bibinfo{author}{M.~Hassanaly}, \bibinfo{author}{P.~J. Weddle}, \bibinfo{author}{R.~N. King}, \bibinfo{author}{S.~De}, \bibinfo{author}{A.~Doostan}, \bibinfo{author}{C.~R. Randall}, \bibinfo{author}{E.~J. Dufek}, \bibinfo{author}{A.~M. Colclasure}, \bibinfo{author}{K.~Smith},
\newblock \bibinfo{title}{Pinn surrogate of li-ion battery models for parameter inference, part ii: Regularization and application of the pseudo-2d model},
\newblock \bibinfo{journal}{Journal of Energy Storage} \bibinfo{volume}{98} (\bibinfo{year}{2024}) \bibinfo{pages}{113104}.
\bibitem[{Smith et~al.(2021)Smith, Smith, Diaz-Ibarra, Parra-Álvarez, Thornock, Spinti, Harding, Marshall, and Fischer}]{smith2021Atikokan}
\bibinfo{author}{P.~J. Smith}, \bibinfo{author}{S.~T. Smith}, \bibinfo{author}{O.~H. Diaz-Ibarra}, \bibinfo{author}{J.~C. Parra-Álvarez}, \bibinfo{author}{J.~N. Thornock}, \bibinfo{author}{J.~Spinti}, \bibinfo{author}{S.~Harding}, \bibinfo{author}{L.~Marshall}, \bibinfo{author}{B.~Fischer},
\newblock \bibinfo{title}{{The Atikokan Digital Twin: Bayesian Physics-based Machine Learning for Low-load Firing in the Atikokan Biomass Utility Boiler}},
\newblock in: \bibinfo{booktitle}{International Journal of Energy for a Clean Environment}, volume~\bibinfo{volume}{24}, \bibinfo{year}{2021}, pp. \bibinfo{pages}{63--78}.
\bibitem[{Vaidheeswaran et~al.(2021)Vaidheeswaran, Gel, Clarke, and Rogers}]{vaidheeswaran2021sensitivity}
\bibinfo{author}{A.~Vaidheeswaran}, \bibinfo{author}{A.~Gel}, \bibinfo{author}{M.~A. Clarke}, \bibinfo{author}{W.~Rogers}, \bibinfo{title}{Sensitivity analysis of particle-in-cell modeling parameters in settling bed, bubbling fluidized bed and circulating fluidized bed}, \bibinfo{type}{Technical Report}, National Energy Technology Laboratory (NETL), Pittsburgh, PA, Morgantown, WV~…, \bibinfo{year}{2021}.
\bibitem[{Gao et~al.(2021)Gao, Zhang, Zhu, and Guo}]{gao2021global}
\bibinfo{author}{Y.~Gao}, \bibinfo{author}{X.~Zhang}, \bibinfo{author}{C.~Zhu}, \bibinfo{author}{B.~Guo},
\newblock \bibinfo{title}{Global parameter sensitivity analysis of electrochemical model for lithium-ion batteries considering aging},
\newblock \bibinfo{journal}{IEEE/ASME Transactions on Mechatronics} \bibinfo{volume}{26} (\bibinfo{year}{2021}) \bibinfo{pages}{1283--1294}.
\bibitem[{Volger et~al.(2023)Volger, Puiman, and Haringa}]{volger2023bubbles}
\bibinfo{author}{R.~Volger}, \bibinfo{author}{L.~Puiman}, \bibinfo{author}{C.~Haringa},
\newblock \bibinfo{title}{{Bubbles and Broth: A review on the impact of broth composition on bubble column bioreactor hydrodynamics}},
\newblock \bibinfo{journal}{Biochemical Engineering Journal}  (\bibinfo{year}{2023}) \bibinfo{pages}{109124}.
\bibitem[{McClure et~al.(2015)McClure, Lee, Kavanagh, Fletcher, and Barton}]{mcclure2015impact}
\bibinfo{author}{D.~D. McClure}, \bibinfo{author}{A.~C. Lee}, \bibinfo{author}{J.~M. Kavanagh}, \bibinfo{author}{D.~F. Fletcher}, \bibinfo{author}{G.~W. Barton},
\newblock \bibinfo{title}{Impact of surfactant addition on oxygen mass transfer in a bubble column},
\newblock \bibinfo{journal}{Chemical Engineering \& Technology} \bibinfo{volume}{38} (\bibinfo{year}{2015}) \bibinfo{pages}{44--52}.
\bibitem[{Painmanakul et~al.(2005)Painmanakul, Loubi{\`e}re, H{\'e}brard, Mietton-Peuchot, and Roustan}]{painmanakul2005effect}
\bibinfo{author}{P.~Painmanakul}, \bibinfo{author}{K.~Loubi{\`e}re}, \bibinfo{author}{G.~H{\'e}brard}, \bibinfo{author}{M.~Mietton-Peuchot}, \bibinfo{author}{M.~Roustan},
\newblock \bibinfo{title}{Effect of surfactants on liquid-side mass transfer coefficients},
\newblock \bibinfo{journal}{Chemical Engineering Science} \bibinfo{volume}{60} (\bibinfo{year}{2005}) \bibinfo{pages}{6480--6491}.
\bibitem[{Weller et~al.(1998)Weller, Tabor, Jasak, and Fureby}]{weller1998}
\bibinfo{author}{H.~G. Weller}, \bibinfo{author}{G.~Tabor}, \bibinfo{author}{H.~Jasak}, \bibinfo{author}{C.~Fureby},
\newblock \bibinfo{title}{A tensorial approach to computational continuum mechanics using object-oriented techniques},
\newblock \bibinfo{journal}{Computers in Physics} \bibinfo{volume}{12} (\bibinfo{year}{1998}) \bibinfo{pages}{620--631}.
\bibitem[{Rahimi et~al.(2018)Rahimi, Sitaraman, Humbird, and Stickel}]{rahimi2018}
\bibinfo{author}{M.~J. Rahimi}, \bibinfo{author}{H.~Sitaraman}, \bibinfo{author}{D.~Humbird}, \bibinfo{author}{J.~J. Stickel},
\newblock \bibinfo{title}{{Computational fluid dynamics study of full-scale aerobic bioreactors: Evaluation of gas--liquid mass transfer, oxygen uptake, and dynamic oxygen distribution}},
\newblock \bibinfo{journal}{Chemical Engineering Research and Design} \bibinfo{volume}{139} (\bibinfo{year}{2018}) \bibinfo{pages}{283--295}.
\bibitem[{Lehnigk et~al.(2022)Lehnigk, Bainbridge, Liao, Lucas, Niemi, Peltola, and Schlegel}]{lehnigk2022open}
\bibinfo{author}{R.~Lehnigk}, \bibinfo{author}{W.~Bainbridge}, \bibinfo{author}{Y.~Liao}, \bibinfo{author}{D.~Lucas}, \bibinfo{author}{T.~Niemi}, \bibinfo{author}{J.~Peltola}, \bibinfo{author}{F.~Schlegel},
\newblock \bibinfo{title}{An open-source population balance modeling framework for the simulation of polydisperse multiphase flows},
\newblock \bibinfo{journal}{AIChE Journal} \bibinfo{volume}{68} (\bibinfo{year}{2022}) \bibinfo{pages}{e17539}.
\bibitem[{Sitaraman et~al.(2023)Sitaraman, Lischeske, Lu, and Stickel}]{sitaraman2023reacting}
\bibinfo{author}{H.~Sitaraman}, \bibinfo{author}{J.~Lischeske}, \bibinfo{author}{Y.~Lu}, \bibinfo{author}{J.~Stickel},
\newblock \bibinfo{title}{A reacting multiphase computational flow model for 2 3-butanediol synthesis in industrial-scale bioreactors},
\newblock \bibinfo{journal}{Chemical Engineering Research and Design} \bibinfo{volume}{197} (\bibinfo{year}{2023}) \bibinfo{pages}{38--52}.
\bibitem[{Passalacqua and Fox(2011)}]{passalacqua2011implementation}
\bibinfo{author}{A.~Passalacqua}, \bibinfo{author}{R.~O. Fox},
\newblock \bibinfo{title}{Implementation of an iterative solution procedure for multi-fluid gas--particle flow models on unstructured grids},
\newblock \bibinfo{journal}{Powder Technology} \bibinfo{volume}{213} (\bibinfo{year}{2011}) \bibinfo{pages}{174--187}.
\bibitem[{Grace(1976)}]{grace1976shapes}
\bibinfo{author}{J.~R. Grace},
\newblock \bibinfo{title}{{Shapes and Velocities of Single Drops and Bubbles Moving Freely through Immisicible Liquids}},
\newblock \bibinfo{journal}{Trans. Inst. Chem. Eng.} \bibinfo{volume}{54} (\bibinfo{year}{1976}) \bibinfo{pages}{167--173}.
\bibitem[{Tomiyama et~al.(2002)Tomiyama, Tamai, Zun, and Hosokawa}]{tomiyama2002transverse}
\bibinfo{author}{A.~Tomiyama}, \bibinfo{author}{H.~Tamai}, \bibinfo{author}{I.~Zun}, \bibinfo{author}{S.~Hosokawa},
\newblock \bibinfo{title}{Transverse migration of single bubbles in simple shear flows},
\newblock \bibinfo{journal}{Chemical Engineering Science} \bibinfo{volume}{57} (\bibinfo{year}{2002}) \bibinfo{pages}{1849--1858}.
\bibitem[{Antal et~al.(1991)Antal, Lahey~Jr, and Flaherty}]{antal1991analysis}
\bibinfo{author}{S.~Antal}, \bibinfo{author}{R.~Lahey~Jr}, \bibinfo{author}{J.~Flaherty},
\newblock \bibinfo{title}{Analysis of phase distribution in fully developed laminar bubbly two-phase flow},
\newblock \bibinfo{journal}{International Journal of Multiphase Flow} \bibinfo{volume}{17} (\bibinfo{year}{1991}) \bibinfo{pages}{635--652}.
\bibitem[{Burns et~al.(2004)Burns, Frank, Hamill, and Shi}]{burns2004favre}
\bibinfo{author}{A.~D. Burns}, \bibinfo{author}{T.~Frank}, \bibinfo{author}{I.~Hamill}, \bibinfo{author}{J.-M. Shi},
\newblock \bibinfo{title}{{The Favre averaged drag model for turbulent dispersion in Eulerian multi-phase flows}},
\newblock in: \bibinfo{booktitle}{5th International Conference on Multiphase Flow, ICMF}, volume~\bibinfo{volume}{4}, \bibinfo{organization}{ICMF}, \bibinfo{year}{2004}, pp. \bibinfo{pages}{1--17}.
\bibitem[{Higbie(1935)}]{higbie1935rate}
\bibinfo{author}{R.~Higbie},
\newblock \bibinfo{title}{The rate of absorption of pure gas into a still liquid during short periods of exposure},
\newblock \bibinfo{journal}{Trans. Am. Inst. Chem. Engrs.} \bibinfo{volume}{31} (\bibinfo{year}{1935}) \bibinfo{pages}{365--389}.
\bibitem[{Behzadi et~al.(2004)Behzadi, Issa, and Rusche}]{behzadi2004modelling}
\bibinfo{author}{A.~Behzadi}, \bibinfo{author}{R.~Issa}, \bibinfo{author}{H.~Rusche},
\newblock \bibinfo{title}{Modelling of dispersed bubble and droplet flow at high phase fractions},
\newblock \bibinfo{journal}{Chemical Engineering Science} \bibinfo{volume}{59} (\bibinfo{year}{2004}) \bibinfo{pages}{759--770}.
\bibitem[{Lahey~Jr(2005)}]{lahey2005simulation}
\bibinfo{author}{R.~T. Lahey~Jr},
\newblock \bibinfo{title}{The simulation of multidimensional multiphase flows},
\newblock \bibinfo{journal}{Nuclear Engineering and Design} \bibinfo{volume}{235} (\bibinfo{year}{2005}) \bibinfo{pages}{1043--1060}.
\bibitem[{Popovac and Hanjalic(2007)}]{popovac2007compound}
\bibinfo{author}{M.~Popovac}, \bibinfo{author}{K.~Hanjalic},
\newblock \bibinfo{title}{{Compound wall treatment for RANS computation of complex turbulent flows and heat transfer}},
\newblock \bibinfo{journal}{Flow, Turbulence and Combustion} \bibinfo{volume}{78} (\bibinfo{year}{2007}) \bibinfo{pages}{177--202}.
\bibitem[{Buwa and Ranade(2002)}]{buwa2002dynamics}
\bibinfo{author}{V.~V. Buwa}, \bibinfo{author}{V.~V. Ranade},
\newblock \bibinfo{title}{{Dynamics of gas--liquid flow in a rectangular bubble column: experiments and single/multi-group CFD simulations}},
\newblock \bibinfo{journal}{Chemical Engineering Science} \bibinfo{volume}{57} (\bibinfo{year}{2002}) \bibinfo{pages}{4715--4736}.
\bibitem[{D{\'\i}az et~al.(2008)D{\'\i}az, Iranzo, Cuadra, Barbero, Montes, and Gal{\'a}n}]{diaz2008numerical}
\bibinfo{author}{M.~E. D{\'\i}az}, \bibinfo{author}{A.~Iranzo}, \bibinfo{author}{D.~Cuadra}, \bibinfo{author}{R.~Barbero}, \bibinfo{author}{F.~J. Montes}, \bibinfo{author}{M.~A. Gal{\'a}n},
\newblock \bibinfo{title}{Numerical simulation of the gas--liquid flow in a laboratory scale bubble column: influence of bubble size distribution and non-drag forces},
\newblock \bibinfo{journal}{Chemical Engineering Journal} \bibinfo{volume}{139} (\bibinfo{year}{2008}) \bibinfo{pages}{363--379}.
\bibitem[{Colella et~al.(1999)Colella, Vinci, Bagatin, Masi, and Bakr}]{colella1999study}
\bibinfo{author}{D.~Colella}, \bibinfo{author}{D.~Vinci}, \bibinfo{author}{R.~Bagatin}, \bibinfo{author}{M.~Masi}, \bibinfo{author}{E.~A. Bakr},
\newblock \bibinfo{title}{A study on coalescence and breakage mechanisms in three different bubble columns},
\newblock \bibinfo{journal}{Chemical Engineering Science} \bibinfo{volume}{54} (\bibinfo{year}{1999}) \bibinfo{pages}{4767--4777}.
\bibitem[{Lucas and Ziegenhein(2019)}]{lucas2019influence}
\bibinfo{author}{D.~Lucas}, \bibinfo{author}{T.~Ziegenhein},
\newblock \bibinfo{title}{Influence of the bubble size distribution on the bubble column flow regime},
\newblock \bibinfo{journal}{International Journal of Multiphase Flow} \bibinfo{volume}{120} (\bibinfo{year}{2019}) \bibinfo{pages}{103092}.
\bibitem[{Ramezani et~al.(2012)Ramezani, Mostoufi, and Mehrnia}]{ramezani2012improved}
\bibinfo{author}{M.~Ramezani}, \bibinfo{author}{N.~Mostoufi}, \bibinfo{author}{M.~R. Mehrnia},
\newblock \bibinfo{title}{Improved modeling of bubble column reactors by considering the bubble size distribution},
\newblock \bibinfo{journal}{Industrial \& Engineering Chemistry Research} \bibinfo{volume}{51} (\bibinfo{year}{2012}) \bibinfo{pages}{5705--5714}.
\bibitem[{Kumar and Ramkrishna(1996)}]{kumar1996solution}
\bibinfo{author}{S.~Kumar}, \bibinfo{author}{D.~Ramkrishna},
\newblock \bibinfo{title}{{On the solution of population balance equations by discretization—I. A fixed pivot technique}},
\newblock \bibinfo{journal}{Chemical Engineering Science} \bibinfo{volume}{51} (\bibinfo{year}{1996}) \bibinfo{pages}{1311--1332}.
\bibitem[{Liao et~al.(2018)Liao, Oertel, Kriebitzsch, Schlegel, and Lucas}]{liao2018discrete}
\bibinfo{author}{Y.~Liao}, \bibinfo{author}{R.~Oertel}, \bibinfo{author}{S.~Kriebitzsch}, \bibinfo{author}{F.~Schlegel}, \bibinfo{author}{D.~Lucas},
\newblock \bibinfo{title}{A discrete population balance equation for binary breakage},
\newblock \bibinfo{journal}{International Journal for Numerical Methods in Fluids} \bibinfo{volume}{87} (\bibinfo{year}{2018}) \bibinfo{pages}{202--215}.
\bibitem[{Lehr et~al.(2002)Lehr, Millies, and Mewes}]{lehr2002bubble}
\bibinfo{author}{F.~Lehr}, \bibinfo{author}{M.~Millies}, \bibinfo{author}{D.~Mewes},
\newblock \bibinfo{title}{Bubble-size distributions and flow fields in bubble columns},
\newblock \bibinfo{journal}{AIChE Journal} \bibinfo{volume}{48} (\bibinfo{year}{2002}) \bibinfo{pages}{2426--2443}.
\bibitem[{Kostoglou and Karabelas(2005)}]{kostoglou2005toward}
\bibinfo{author}{M.~Kostoglou}, \bibinfo{author}{A.~Karabelas},
\newblock \bibinfo{title}{Toward a unified framework for the derivation of breakage functions based on the statistical theory of turbulence},
\newblock \bibinfo{journal}{Chemical Engineering Science} \bibinfo{volume}{60} (\bibinfo{year}{2005}) \bibinfo{pages}{6584--6595}.
\bibitem[{Hissanaga et~al.(2020)Hissanaga, Padoin, and Paladino}]{hissanaga2020mass}
\bibinfo{author}{A.~Hissanaga}, \bibinfo{author}{N.~Padoin}, \bibinfo{author}{E.~Paladino},
\newblock \bibinfo{title}{Mass transfer modeling and simulation of a transient homogeneous bubbly flow in a bubble column},
\newblock \bibinfo{journal}{Chemical Engineering Science} \bibinfo{volume}{218} (\bibinfo{year}{2020}) \bibinfo{pages}{115531}.
\bibitem[{Ngu et~al.(2022)Ngu, Morchain, and Cockx}]{ngu2022spatio}
\bibinfo{author}{V.~Ngu}, \bibinfo{author}{J.~Morchain}, \bibinfo{author}{A.~Cockx},
\newblock \bibinfo{title}{{Spatio-temporal 1D gas--liquid model for biological methanation in lab scale and industrial bubble column}},
\newblock \bibinfo{journal}{Chemical Engineering Science} \bibinfo{volume}{251} (\bibinfo{year}{2022}) \bibinfo{pages}{117478}.
\bibitem[{Rzehak and Krepper(2016)}]{rzehak2016euler}
\bibinfo{author}{R.~Rzehak}, \bibinfo{author}{E.~Krepper},
\newblock \bibinfo{title}{{Euler-Euler simulation of mass-transfer in bubbly flows}},
\newblock \bibinfo{journal}{Chemical Engineering Science} \bibinfo{volume}{155} (\bibinfo{year}{2016}) \bibinfo{pages}{459--468}.
\bibitem[{Huang et~al.(2018)Huang, McClure, Barton, Fletcher, and Kavanagh}]{huang2018assessment}
\bibinfo{author}{Z.~Huang}, \bibinfo{author}{D.~D. McClure}, \bibinfo{author}{G.~W. Barton}, \bibinfo{author}{D.~F. Fletcher}, \bibinfo{author}{J.~M. Kavanagh},
\newblock \bibinfo{title}{Assessment of the impact of bubble size modelling in cfd simulations of alternative bubble column configurations operating in the heterogeneous regime},
\newblock \bibinfo{journal}{Chemical Engineering Science} \bibinfo{volume}{186} (\bibinfo{year}{2018}) \bibinfo{pages}{88--101}.
\bibitem[{Yue et~al.(2007)Yue, Chen, Yuan, Luo, and Gonthier}]{yue2007hydrodynamics}
\bibinfo{author}{J.~Yue}, \bibinfo{author}{G.~Chen}, \bibinfo{author}{Q.~Yuan}, \bibinfo{author}{L.~Luo}, \bibinfo{author}{Y.~Gonthier},
\newblock \bibinfo{title}{Hydrodynamics and mass transfer characteristics in gas--liquid flow through a rectangular microchannel},
\newblock \bibinfo{journal}{Chemical Engineering Science} \bibinfo{volume}{62} (\bibinfo{year}{2007}) \bibinfo{pages}{2096--2108}.
\bibitem[{Mueller and Raman(2018)}]{mueller2018model}
\bibinfo{author}{M.~E. Mueller}, \bibinfo{author}{V.~Raman},
\newblock \bibinfo{title}{{Model form uncertainty quantification in turbulent combustion simulations: Peer models}},
\newblock \bibinfo{journal}{Combustion and Flame} \bibinfo{volume}{187} (\bibinfo{year}{2018}) \bibinfo{pages}{137--146}.
\bibitem[{Hassanaly et~al.(2024{\natexlab{a}})Hassanaly, Sitaraman, Rahimi, and Parra-Alvarez}]{birdswr}
\bibinfo{author}{M.~Hassanaly}, \bibinfo{author}{H.~Sitaraman}, \bibinfo{author}{M.~Rahimi}, \bibinfo{author}{M.~Parra-Alvarez}, \bibinfo{title}{{BiRD (BioReactorDesign) [SWR-24-35]}}, \bibinfo{year}{2024}{\natexlab{a}}. \URLprefix \url{https://www.osti.gov/biblio/2319227}. \DOIprefix\doi{10.11578/dc.20240307.3}.
\bibitem[{Hassanaly et~al.(2024{\natexlab{b}})Hassanaly, Weddle, King, De, Doostan, Randall, Dufek, Colclasure, and Smith}]{hassanaly2023pinn1}
\bibinfo{author}{M.~Hassanaly}, \bibinfo{author}{P.~J. Weddle}, \bibinfo{author}{R.~N. King}, \bibinfo{author}{S.~De}, \bibinfo{author}{A.~Doostan}, \bibinfo{author}{C.~R. Randall}, \bibinfo{author}{E.~J. Dufek}, \bibinfo{author}{A.~M. Colclasure}, \bibinfo{author}{K.~Smith},
\newblock \bibinfo{title}{Pinn surrogate of li-ion battery models for parameter inference, part i: Implementation and multi-fidelity hierarchies for the single-particle model},
\newblock \bibinfo{journal}{Journal of Energy Storage} \bibinfo{volume}{98} (\bibinfo{year}{2024}{\natexlab{b}}) \bibinfo{pages}{113103}.
\bibitem[{Khalil et~al.(2015)Khalil, Lacaze, Oefelein, and Najm}]{khalil2015uncertainty}
\bibinfo{author}{M.~Khalil}, \bibinfo{author}{G.~Lacaze}, \bibinfo{author}{J.~C. Oefelein}, \bibinfo{author}{H.~N. Najm},
\newblock \bibinfo{title}{{Uncertainty quantification in LES of a turbulent bluff-body stabilized flame}},
\newblock \bibinfo{journal}{Proceedings of the Combustion Institute} \bibinfo{volume}{35} (\bibinfo{year}{2015}) \bibinfo{pages}{1147--1156}.
\bibitem[{Raman and Hassanaly(2019)}]{raman2019emerging}
\bibinfo{author}{V.~Raman}, \bibinfo{author}{M.~Hassanaly},
\newblock \bibinfo{title}{Emerging trends in numerical simulations of combustion systems},
\newblock \bibinfo{journal}{Proceedings of the Combustion Institute} \bibinfo{volume}{37} (\bibinfo{year}{2019}) \bibinfo{pages}{2073--2089}.
\bibitem[{Stagonas et~al.(2011)Stagonas, Warbrick, Muller, and Magagna}]{stagonas2011surface}
\bibinfo{author}{D.~Stagonas}, \bibinfo{author}{D.~Warbrick}, \bibinfo{author}{G.~Muller}, \bibinfo{author}{D.~Magagna},
\newblock \bibinfo{title}{Surface tension effects on energy dissipation by small scale, experimental breaking waves},
\newblock \bibinfo{journal}{Coastal Engineering} \bibinfo{volume}{58} (\bibinfo{year}{2011}) \bibinfo{pages}{826--836}.
\bibitem[{O'Mahony(1972)}]{o1972purity}
\bibinfo{author}{M.~O'Mahony},
\newblock \bibinfo{title}{Purity effects and distilled water taste},
\newblock \bibinfo{journal}{Nature} \bibinfo{volume}{240} (\bibinfo{year}{1972}) \bibinfo{pages}{489--489}.
\bibitem[{Oliver et~al.(2014)Oliver, Malaya, Ulerich, and Moser}]{oliver2014estimating}
\bibinfo{author}{T.~A. Oliver}, \bibinfo{author}{N.~Malaya}, \bibinfo{author}{R.~Ulerich}, \bibinfo{author}{R.~D. Moser},
\newblock \bibinfo{title}{Estimating uncertainties in statistics computed from direct numerical simulation},
\newblock \bibinfo{journal}{Physics of Fluids} \bibinfo{volume}{26} (\bibinfo{year}{2014}).
\bibitem[{Hoffman et~al.(2014)Hoffman, Gelman et~al.}]{hoffman2014no}
\bibinfo{author}{M.~Hoffman}, \bibinfo{author}{A.~Gelman}, et~al.,
\newblock \bibinfo{title}{{The No-U-Turn sampler: adaptively setting path lengths in Hamiltonian Monte Carlo}},
\newblock \bibinfo{journal}{Journal Machine Learning Research} \bibinfo{volume}{15} (\bibinfo{year}{2014}) \bibinfo{pages}{1593--1623}.
\bibitem[{Phan et~al.(2019)Phan, Pradhan, and Jankowiak}]{numpyro}
\bibinfo{author}{D.~Phan}, \bibinfo{author}{N.~Pradhan}, \bibinfo{author}{M.~Jankowiak},
\newblock \bibinfo{title}{{Composable effects for flexible and accelerated probabilistic programming in NumPyro}},
\newblock \bibinfo{journal}{arXiv preprint arXiv:1912.11554}  (\bibinfo{year}{2019}).
\bibitem[{Kvam and Sarkisov(2019)}]{kvam2019solubility}
\bibinfo{author}{O.~Kvam}, \bibinfo{author}{L.~Sarkisov},
\newblock \bibinfo{title}{Solubility prediction in mixed solvents: A combined molecular simulation and experimental approach},
\newblock \bibinfo{journal}{Fluid Phase Equilibria} \bibinfo{volume}{484} (\bibinfo{year}{2019}) \bibinfo{pages}{26--37}.
\bibitem[{Saltelli(2002)}]{saltelli2002making}
\bibinfo{author}{A.~Saltelli},
\newblock \bibinfo{title}{Making best use of model evaluations to compute sensitivity indices},
\newblock \bibinfo{journal}{Computer physics communications} \bibinfo{volume}{145} (\bibinfo{year}{2002}) \bibinfo{pages}{280--297}.

\end{thebibliography}
\end{document}